\documentclass[11pt,a4paper]{article}
\pdfoutput=1

\usepackage{jcappub}
\usepackage{amsfonts}
\usepackage{amsmath}
\usepackage{amssymb}
\usepackage{aas_macros}
\usepackage{graphicx}
\usepackage{color}
\usepackage{float}
\usepackage{bm}
\usepackage{natbib}
\usepackage{hyperref}
\usepackage{longtable}
\usepackage{comment}
\usepackage{subcaption}
\usepackage{wasysym}
\usepackage{mathtools}
\usepackage{fancyhdr}
\usepackage{xparse}
\usepackage{xcolor}
\usepackage{physics}
\usepackage{enumerate} 
\usepackage{tensor}

\def\lsim{~\rlap{$<$}{\lower 1.0ex\hbox{$\sim$}}}
\def\bsim{~\rlap{$>$}{\lower 1.0ex\hbox{$\sim$}}}

\def\kpc{\ {\rm kpc}}
\def\AU{\ {\rm AU}}

\def\msun{\ {\rm M_\odot}}

\def\apjl{Astrophys.\ J.\ Lett.}

\def\aap{Astron.\ Astrophys.}

\def\apjs{Astrophys.\ J. Supp.}

\newcommand{\be}{\begin{equation}}
\newcommand{\ee}{\end{equation}}

\let\ln\relax
\DeclareMathOperator{\ln}{ln}

\def\mathbi#1{\textbf{\em #1}}

\def\vr{\mathbi{r}}

\def\vxi{\boldsymbol{\xi}}

\def\fpbh{f_\text{pbh}}
\def\mpbh{M_\text{pbh}}
\def\as{a_*}
\def\ss{s_*}

\newcommand{\ms}{\mathcal{S}}

\def\emph#1{\textit{#1}}

\def\comment#1{}
\definecolor{RedWine}{rgb}{0.743,0,0}
\definecolor{RoyalBlue}{rgb}{0.25,.41,.88}
\definecolor{ForestGreen}{rgb}{.13,.54,.13}
\definecolor{DeepPurple}{rgb}{.72,.18,1}

\allowdisplaybreaks

\title{Detection prospects for the GW background of Galactic (sub)solar mass primordial black holes}

\author[a]{Frans van Die,}
\author[a]{Ivan Rapoport,}
\author[b,c]{Yonadav Barry Ginat,}
\author[a,d]{and Vincent Desjacques}

\affiliation[a]{Physics department, Technion, 3200003 Haifa, Israel}
\affiliation[b]{Rudolf Peierls Centre for Theoretical Physics, University of Oxford, Parks Road, Oxford, OX1 3PU, United Kingdom}
\affiliation[c]{New College, Holywell Street, Oxford, OX1 3BN, United Kingdom}
\affiliation[d]{Asher Space Research Institute, Technion, 3200003 Haifa, Israel}

\emailAdd{frans.van@campus.technion.ac.il}
\emailAdd{ivanr@campus.technion.ac.il}
\emailAdd{yb.ginat@physics.ox.ac.uk}
\emailAdd{dvince@physics.technion.ac.il}

\date{\today}

\label{firstpage}

\abstract{In multi-component dark matter models, a fraction $f_\text{pbh}$ of the dark matter could be in the form of primordial black holes (PBHs) with (sub)solar masses. Some would have formed binaries that presently trace the Milky Way halo of particle dark matter. We explore the gravitational wave (GW) signal produced by such a hypothetical population of Galactic PBH binaries and assess its detectability by the LISA experiment. 
For this purpose, we model the formation and evolution of early-type PBH binaries accounting for GW hardening and binary disruption in the Milky Way. Our analysis reveals that the present-day Galactic population of PBH binaries is characterized by very high orbital eccentricities $|1-e|\ll 1$. For a PBH mass $M_{\rm pbh} \sim 0.1 - 1 M_\odot$, this yields a GW background that peaks in the millihertz frequency range where the LISA instrumental noise is minimum. While this signal remains below the LISA detection threshold for viable $f_\text{pbh}\lesssim 0.01$, future GW observatories such as DECIGO and BBO could detect it if $0.01\lesssim M_{\rm pbh} \lesssim 0.1 M_\odot$. Furthermore, we anticipate that, after 5 years of observations, LISA should be able to detect $\mathcal{O}(100)$ (resp. $\mathcal{O}(1)$) loud Galactic PBH binaries of mass $M_{\rm pbh} \sim 0.1 - 1 M_\odot$ with a SNR $\geq 5$ if $f_{\rm pbh}=0.01$ (resp. $f_{\rm pbh}=0.001$). Nonlinear effects not considered here such as mass accretion and dynamical capture could alter these predictions.}

\begin{document}

\maketitle

\section{Introduction}
\label{sec:intro}

Primordial black holes (PBHs) could have formed in the early Universe as a result of high-density fluctuations \cite{1971MNRAS.152...75H, Carr:1975qj,ivanov/naselsky/novikov:1994,bellido/etal:1996,ivanov:1998}. Interest in PBHs has surged since the LIGO-Virgo-KAGRA (LVK) experiments unveiled a population of black holes (BHs) with characteristic masses comparable or larger than the stellar-mass black hole candidates located in the Milky Way (MW) \cite{Abbott_2016, Abbott_2019, Abbott_2021}, which led to the suggestion that some of the black holes binaries detected by GW experiments are primordial \cite{Bird_2016,blinnikov/etal:2016}.

In contrast to earlier models where all dark matter (DM) was thought to consist of PBHs, it is now believed that PBHs could span a wide range of masses and represent only a fraction of the total energy density of a multi-component dark sector (for reviews, see \cite{sasaki/etal:2018, Carr_2020} and references therein). The PBH abundance, as parameterized by the fraction of DM in the form of PBHs is written as
\begin{equation}
    \fpbh = \frac{\Omega_{\text{pbh}}}{\Omega_{\text{DM}}}
    \label{eq:fpbh}
\end{equation}
Current observational constraints on PBH mass and abundance (summarized in \cite{Carr_2021, Villanueva_Domingo_2021}) based on the microlensing of distant supernovae \cite{Zumalac_rregui_2018,derocco2023rogue, Mroz:2024mse,Mediavilla:2024yhh}, cosmic microwave background (CMB) anisotropies \cite{ricotti/etal:2008,aloni/etal:2017}, gravimeter analyses \cite{bertrand2023observing,Cuadrat-Grzybowski:2024uph}, observed GW events \citep[e.g.][]{Andr_s_Carcasona_2024, Huang:2024wse} 
and interactions with compact stellar remnants \cite{tran2023close,yamamoto2023prospects} yield an upper bound $\fpbh \lesssim 0.01$ for PBH in the mass window  $10^{-2} - 10^3\msun$, with slightly tighter constraints at the high end of this mass window. Enhanced microlensing constraints with ROMAN could probe PBH fractions down to $\fpbh\sim 10^{-3}-10^{-4}$ and may also investigate asteroid-mass PBHs \cite{Fardeen:2023euf,derocco2023rogue}, which remain largely unconstrained \cite{Dent:2024yje}. While most constraints are derived from a monochromatic mass function, it is important to note that introducing extended mass functions can somewhat modify these constraints \citep[e.g.][]{Dizon:2024iao}.

Close pairs of PBHs can form binaries in the early Universe and emit GWs at frequencies dependent on their masses (for reviews, see \cite{domènech2024probing,domènech2023lectures, LISACosmologyWorkingGroup:2023njw}). They can merge by the present epoch if their initial eccentricity is close to unity \cite{nakamura/etal:1997, Bird_2016,sasaki/etal:2016, Ali_Ha_moud_2017,raidal/etal:2024}. However, most of the PBH binaries that ever form will not have had enough time to have merged by now. These can be disrupted, or further harden through interactions with astrophysical objects or other PBHs.
For low PBH fractions $\fpbh \lesssim 0.01$, consistent with current observational limits, PBH binaries get advected onto the DM halos \cite{mack/etal:2007, rice/zhang:2017,inman/alihaimoud:2019} and masquerade as stellar BH binaries (some of which could reside in globular clusters, see \cite{Vanzan:2024wwc}). In the MW halo, they would contribute to the Galactic GW background, which arises due to the superposition of many GW signals and can be probed by future GW space experiments such as the Laser Interferometer Space Antenna (LISA).
Other Galactic sources of GWs include stellar compact binaries, to which $\sim 10^8$ double white dwarfs (DWDs) are the dominant contribution \cite{Maoz:2016bxg,Strub:2024kbe}. 

The number $N_0$ of PBH binaries expected to reside in the MW halo is given by
\begin{equation}
    N_0 =  \frac{M_\text{DM}}{\text{M}_{\odot}}\frac{\fpbh \eta_0}{2m } \simeq 1.18\times 10^{10}\, m^{-1}\left(\frac{f}{0.01}\right)\, \eta_0
    \label{numberpbhbs}
\end{equation}
assuming a MW DM mass of $M_\text{DM}= 2\times 10^{12} \msun$ \cite{phelps/etal:2013,Grand:2019rma} and a monochromatic PBH mass spectrum with $\mpbh=m \msun$. For convenience, we have introduced the fraction $f = 0.85\fpbh$ of all matter in PBHs, while $\eta_0$ represents the fraction of Galactic PBHs that exist in a binary system at the present epoch. As we shall see later, $\eta_0 = \eta_0(f)$ is a function dependent mostly on $f$. For a PBH binary fraction of $\eta_0 \sim 0.1$ as advocated by \cite{Raidal_2019} (which is fortuitously close to the fraction of white dwarfs in DWDs, see \cite{Maoz:2016bxg}) and for the parameter values $f\sim 0.01$ and $m\sim 1$, we find $N_0\sim 10^9$ comparable to the number of DWDs in the MW.  
This motivates our calculation of the GW signal produced by a hypothetical population of Galactic PBH binaries. As we will see below, however, the spatial distribution and orbital parameters of Galactic PBH binaries differ significantly from those of the Galactic DWD population, even when the PBH mass falls in the solar mass range. 

For the sake of generality, we remain agnostic about the precise formation mechanisms of PBHs and consider a multi-component DM scenario, in which a monochromatic population of PBHs of mass in the range $0.01\leq m \leq 1000$ formed at the end of inflation and make up a fraction $f\leq 0.01$ of the total matter\footnote{See e.g. \cite{Young:2019gfc, Mukherjee:2021itf, Liu:2018ess} for studies involving a non-monochromatic PBH mass distribution.}. We will be particularly interested in (sub)solar mass PBHs. Subsolar mass compact objects almost surely have a primordial origin \cite{Carr:1975qj}, as standard stellar evolution models predict that stellar black holes should have masses exceeding solar \cite{burrows/vartanya:2021}, with the caveat that neutron stars and white dwarfs could theoretically be mildly subsolar in mass \cite{shapiro/teukolsky:1983,kilic/etal:2007,metzger/etal:2024}. In the LVK O3 run, three subsolar mass compact object candidates were identified, though not confirmed due to their high false alarm rate \cite{Abbott_2022,Prunier:2024eed}.

In this study, we concentrate solely on PBH binaries forming in the early Universe and discard the possibility that they interact with ordinary astrophysical objects to create mixed primordial-stellar binaries or triples \cite{yamamoto2023prospects,Bhalla:2024jbu}. Therefore, the PBH binary population we are interested in is characterized by very high eccentricities. This distinguishes them from stellar BH binaries or PBH binaries forming in the late Universe via dynamical capture. 

The paper is organized as follows. 
Sections \S\ref{sec:PBHformation} and \S\ref{sec:PBHevolution} focus on the formation of "early-type" PBH binaries from the RD era until the onset of nonlinear structure formation, and their survival in the MW halo until the present epoch. We don't consider binaries forming during nonlinear structure formation, as these are expected to be less distinguishable from astrophysical binary populations \cite{Delos:2024poq}. The present-day space distributions of Galactic PBH binaries and DWDs are modeled in Section \S\ref{sec:galdist}. All this provides the seeds for the computation of the Galactic GW background and the extraction of loud Galactic PBH binaries discussed in Section \S\ref{sec:GWB}. We conclude in \S\ref{sec:conclusions}.

\section{Formation of PBH binaries}
\label{sec:PBHformation}

We initialize the PBH distribution at the end of inflation as a spatial Poisson process. We thus neglect spatial correlations in the initial clustering of PBHs, which depend heavily on the mechanism leading to PBH formation \citep[e.g.][]{alihaimoud:2018, Desjacques:2018wuu,ballesteros/etal:2018, Stasenko:2024pzd, Young:2019gfc, Bringmann:2018mxj, Mukherjee:2021itf,Belotsky:2018wph}. We also ignore the possibility that PBHs inherit a non-zero spin at formation \cite{deluca/etal:2019}.

For convenience, let $s=a/a_\text{eq}$ be the scale factor normalized to unity at matter-radiation (MR) equality. Throughout this section, we define $s-$comoving scales as equal to physical scales at MR equality, in contrast to regular comoving scales defined w.r.t. the present epoch.
Let $x_i$ represent the initial $s-$comoving distance (at the end of inflation) between nearest PBH neighbors, referred to as 'PBH pairs'. The probability distribution $P(x_i)$ is given by 
\begin{equation}
    P(x_i) = 4\pi n x_i^2 \exp\left(-\frac{4\pi n x_i^3}{3}\right) \;.
\end{equation}
Here $n$ is the $s-$comoving number density of PBHs,
\begin{equation}
    n = \frac{f\rho_\text{eq}}{\mpbh}
\end{equation}
with
\begin{equation}
    \rho_\text{eq} = \frac{\Omega_m}{{a_\text{eq}}^3} \frac{3{H_0}^2}{8\pi G} \simeq 1.7 \cdot 10^{-13}\msun \text{AU}^{-3}
\end{equation}
given as the matter density at equality. For the range of $f$ and $m$ considered here, typical values of $x_i$ are much lower than the $s-$comoving Hubble radius $\mathcal{H}_\text{eq}^{-1}\simeq 6\cdot 10^9\AU$, so that we can compute the evolution of the distance to the nearest neighbor using a Newtonian approximation, as in \cite{Ali_Ha_moud_2017}. 

\subsection{Equation of motion}
\label{sec:EOM}

We model the time evolution of the proper separation $r$ between nearest neighbor PBHs with
\begin{equation}
    \ddot{r} - (\dot{H}+H^2)r+\frac{2G\mpbh}{r^2}\frac{r}{\abs{r}}(1-j^2)=0 \;,
    \label{eq:Newton}
\end{equation}
where dots indicate differentiation with respect to cosmic time, $j\equiv j(t)$ represents the dimensionless or reduced angular momentum of the system,
\begin{equation}
    j \equiv \frac{L}{\mpbh\sqrt{2G\mpbh r}}\;.
\end{equation}
and the Hubble rate (in a MR universe) is given by
\begin{equation}
    H(s) = \bigg(\frac{8\pi G}{3}\rho_\text{eq}\bigg)^{1/2}h(s)\quad \mbox{with}\qquad h(s) = \sqrt{s^{-3}+s^{-4}},
\end{equation}

A PBH pair forms a binary system at the time where $\dot{r} = 0$ for the first time, i.e. when the pair decouples from the Hubble flow. The newly-formed binary has properties that will, here and henceforth, be labeled with a $*$ subscript. In particular, the eccentricity of the binary at formation time follows from
\begin{equation}
    j_* = \sqrt{1-e_*^2}
    \label{angularmomentumeccentricity}
\end{equation}
A bound eccentric binary has $0\leq e_*<1$ by definition.

In this effective one-body approximation, it is essential to take into account the angular-momentum barrier, which leads to a large centrifugal force $F_j$ for non-negligible $j$,
\begin{equation}
    F_j = \frac{L^2}{\mpbh r^3}=\frac{2G\mpbh^2j^2}{r^2} \;.
\end{equation}
We immediately see in Eq. \eqref{eq:Newton} that $j>1$ prevents the PBH pair from ever decoupling from the Hubble flow as they will be on a hyperbolic orbit. As we will show shortly, the time dependence of $j$ can prevent binary formation, even for initial values $j_i$ significantly lower than unity, depending on the initial PBH density. For PBH binaries forming in the radiation-dominated (RD) era, eccentricities are very large ($|1-e|\ll 1$) and the angular-momentum barrier can be safely neglected. It is, however, relevant for PBH binary formation in the matter-dominated (MD) era, where angular momentum grows with time. The next subsection is therefore dedicated to quantifying the physics of the angular-momentum barrier.
 
In order to evolve Eq. \eqref{eq:Newton}, it is convenient to introduce the time-independent variable 
\begin{equation}
    \lambda \equiv \frac{4\pi\rho_\text{eq}x_i^3}{3\mpbh}\;
    \label{lambdadef}
\end{equation}
as well as the reduced separation $X \equiv r/(\lambda x_i)$ and the variable $\ms \equiv s/\lambda$ as a proxy for time. This turns Eq.~(\ref{eq:Newton}) into 
\begin{equation}
    X''+ \frac{2+\lambda\ms}{2+2\lambda\ms}\frac{X-X'\ms}{\ms^2}+\frac{\ms^2}{1+\lambda\ms}\frac{1}{\epsilon^2+X^2}\frac{X}{\abs{X}}(1-j^2) =0
    \label{eq:EOM}
\end{equation}
where a prime denotes differentiation with respect to $\ms$ and a small-scale cutoff $\epsilon$ is introduced to regularize the gravitational attraction of the PBHs in the limit $X\to 0$. A value of $\epsilon$ too small restricts the range of $\ms$ that can be probed numerically, whereas a value too large introduces errors in the determination of the reduced scale factor $\ms_*$ at which the PBHs decouple from the Hubble flow to form a binary. 
Eq.~(\ref{eq:EOM}) shows that deep in the RD era, $\ms\ll 1/\lambda$ and the gravitational pull is proportional to $\ms^2$ while, in the MD era, it grows only as $\ms$. Therefore binary formation is suppressed once the RD era ends.
Ref.~\cite{Ali_Ha_moud_2017} were concerned mainly with PBH binaries merging at the present epoch, which form deep in the RD era. However, for the MW relics we are interested in, formation in the MD era is important, especially for low values of $f$ (see Fig.~\ref{fig:cdfs}).

\subsection{Time evolution of the reduced angular momentum}

The angular-momentum barrier depends on the external torques produced by distant PBHs and by fluctuations in the matter distribution \cite{Eroshenko:2016hmn, Hayasaki:2009ug}.
In what follows in later sections, we generally restrict ourselves to PBH scenarios with $f\lesssim 0.01$. Therefore, we can assume that the variations of $j$ caused by distant PBHs and matter density fluctuations are mutually independent. 

We initialize the initial angular momentum of the system by drawing its value $j_i=j_{pbh}+j_\delta$ from the sum of the independent variables $j_{pbh}$ and $j_\delta$ encoding the angular momentum acquired through distant PBHs and matter fluctuations, respectively. This assumption leads to the probability density function (PDF) for $j_i$ being constructed from the convolution
\begin{equation}
    P(j_i) = P(j_{pbh})\ast P(j_\delta)\;.
    \label{Pj}
\end{equation}
Since the variables are independent, the mean and variance of $j_i$ are additive: 
\begin{align}
    \expectationvalue{j_i} &= \expectationvalue{j_{pbh}}+\expectationvalue{j_{\delta}}
    \label{addmean} \\
    \text{Var}(j_i) &= \text{Var}(j_{pbh}) + \text{Var}(j_\delta)
    \label{addvar}
\end{align}
We adopt the distributions $P(j_{pbh})$ and $P(j_\delta)$ given in  \cite{Ali_Ha_moud_2017,Raidal_2019}. Namely, 
\begin{itemize}
    \item For the torques produced by other PBHs, we use
    \begin{equation}
        P(j_{pbh}) = \frac{j_{pbh}}{{j_f}^2{[1+(j_{pbh}/j_f)^2]}^{3/2}}
    \end{equation}
    where $j_f \equiv \frac{1}{2}\lambda f$ is a characteristic (and time-independent) angular momentum. 
    \item For the torques generated by the matter perturbation $\delta_m$ dominated by the smooth or particle dark-matter (PDM) component, we take a Gaussian distribution:
     \begin{equation}
        P(j_\delta) = \frac{j_\delta}{{\sigma_ \delta}^2}e^{-{j_\delta}^2/(2{\sigma_\delta}^2)}
    \end{equation}
    This effect dominates for $f\ll 1$. The Gaussian distribution arises from the superposition of many large-scale linear perturbations. The mean and standard deviation are given by
    \begin{equation}
        {\sigma_\delta}^2 = \frac{3}{10}\sigma_\text{eq}^2\lambda^2 = \text{Var}(j_\delta) \simeq \expval{j_\delta}^2   
        \label{expmatdens}
    \end{equation}
    where $\sigma_\text{eq}^2$ is the variance of linear matter density perturbations in the RD regime on comoving scales of the initial binary separation. Following \cite{Ali_Ha_moud_2017}, we take $\sigma_\text{eq} = 0.005$ for our numerical estimates\footnote{In many scenarios PBH formation is accompanied by an enhancement of the primordial power at small scales \citep[e.g.][]{Ozsoy:2023ryl, Kawasaki:2018daf, Garcia-Bellido:1996mdl, Germani:2018jgr,Carr:2009jm}. As a result, $\sigma_\text{eq}$ might actually be higher. However, since the expected enhancement is strongly model-dependent, we shall ignore this effect.}. The numerical prefactor $3/10$ arises from a directional averaging procedure involving the Gaussian tidal tensor (see Appendix 2 of \cite{Ali_Ha_moud_2017}). 
\end{itemize}
Torques induced by distant PBH and matter fluctuations are suppressed deep in RD, where the Hubble radius is not much greater than the separation of the PBH pair. These torques gradually increase toward MR equality such that, in the MD era, the reduced angular momentum of the system grows in proportion to the scale factor (see below), at least so long as linear cosmological perturbation theory is valid. For simplicity, we have assumed $j(t)=j_i$ constant throughout RD, with the distribution $P(j_i)$ spelled out above. A refined model for $j(t)$ in RD would not have changed our results significantly, since angular momentum accretion is small in the RD era, and PBH binaries forming before equality have high eccentricities $|1-e_*|\lesssim 1$ \cite{Ali_Ha_moud_2017, Raidal_2019}. 

While we assume $j(t) = j_i$ throughout RD, we take into account the time-dependence of $j(t)$ in the MD era arising from the growth of matter perturbations, which dominates for $f\ll 1$. This time dependence is accounted for in the spirit of a mean-field approach, that is, it purely arises through the time dependence of $\expectationvalue{j_\delta}$ via Eq.~\eqref{expmatdens}. To ensure that this also holds in the MD era, we replace the constant $\sigma_{\text{eq}}$ by a redshift-dependent rms variance $\sigma_m$ such that $\expval{j_\delta} \propto \sigma_m$. Here, $\sigma_m$ is the rms variance of matter fluctuations which in principle is to be computed as
\begin{equation}      
    \sigma_m^2(R,z) = \frac{1}{2\pi^2}\int_0^\infty \dd k k^2P_m(k,z)W_R^2(k)
    \label{sigmacomputation}
\end{equation}
with $P_m(k,z)$ is the linear matter power spectrum at redshift $z$ and $W_R(k)$ e.g. a top-hat filter. This expression depends on time through a growth rate squared present in $P_m(k,z)$, and through the (comoving) softening radius $R = (1+z)r(t)$ where $r(t)$ is the time-dependent physical scale of the PBH pair separation. For the $m,f$ values considered here, $R$ is very small relative to characteristic scales such as the Silk-damping scale, being at most of order a parsec deep in MD\footnote{In principle one could also consider the free-streaming length of some WIMP DM particle $\lambda_{fs}$, and then take $R = (1+z)\text{max}[\lambda_{fs}(m_X), r]$. We found this to be irrelevant for WIMPs in the mass range above MeV. Below this mass, speaking of WIMPs loses its meaning.}. For simplicity, we ignore the time dependence of $R$ and take $\sigma_m \simeq s \sigma_\text{eq}$. For PBH binaries forming late in the MD era, $\sigma_m$ can therefore be 1 - 2 orders of magnitude larger than $\sigma_\text{eq}$ until $j$ starts to exceed unity. 

Summarizing, the time dependence of the angular momentum in our mean-field approach is included in the following manner:
\begin{equation}
    \label{eq:jt}
    j(t) = j(s) = j_i + \sqrt{\frac{3}{10}}\lambda \sigma_\text{eq}\,\Xi(s)
\end{equation} 
where 
\begin{equation}
    \Xi(s)\equiv\begin{cases}0 \qquad\qquad\;\; s < 1\\ s -1 \qquad\;\;\; s\geq 1 \end{cases}
\end{equation}
This way, the constant torque from matter fluctuations in the RD era grows linearly with the scale factor after equality.

The initial, reduced angular momentum $j_i$ is to be drawn randomly from a distribution $P(j_i)$, which is to be generated for various choices of $(\lambda,f)$. Transforming $P(j_i)$ to log-space $P(\text{log}j_i)$ allows us to shift a template distribution to the desired range of $j_i$ determined by parameters $\lambda,f$. In practice, we slightly loosen the interpretation of Eq.~\eqref{addvar} such that both the expectation value and the typical angular momentum (the peak of the distribution) are expressed as the sum of the contributions arising from $j_{pbh}$ and $j_\delta$. 

\begin{figure}[h!]
\centering
\includegraphics[width=1\textwidth]{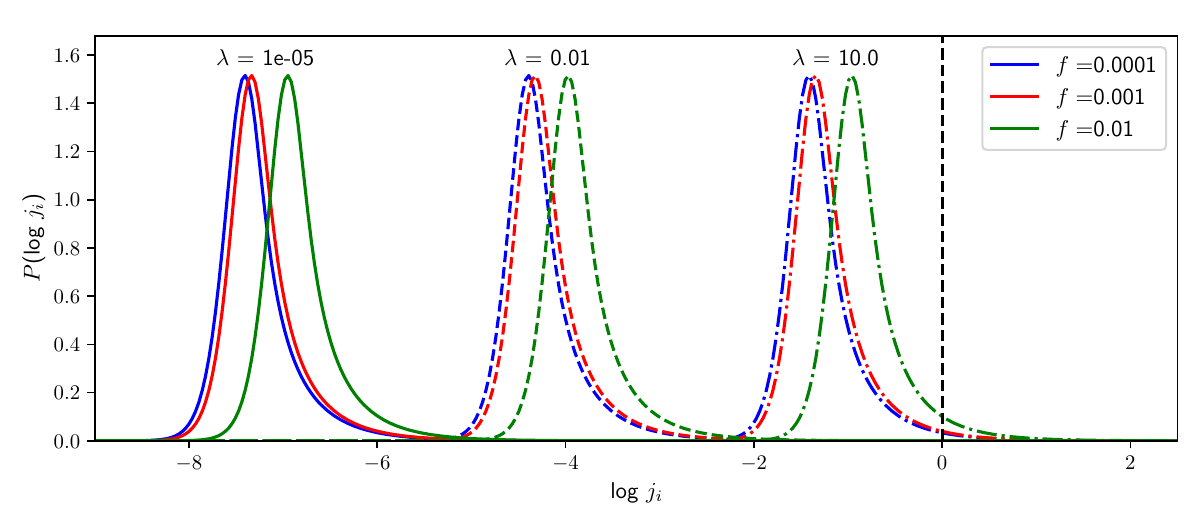}
\caption{The distribution $P(\text{log}j_i)$ of the reduced, initial angular momentum $j_i$ of a PBH pair for different choices of $\lambda$. The solid curves assume the minimal value $\lambda=10^{-5}$ of our parameter space, the dashed lines have $\lambda = 10^{-2}$ and the dashed-dotted distributions assume $\lambda = 10$, which is close to the maximum value of $\lambda$ allowed for binary formation (see Fig. \ref{fig:secondinitial2dpdf}). The black dashed line shows $j_i = 1$, beyond which stable systems never form.}
\label{fig:template_shift}
\end{figure}

Fig.~\ref{fig:template_shift} shows $P(\log j_i)$ for various choices of $\lambda$. The two contributions to the torque have characteristic scales $j_f \sim \lambda f$ (for the other PBHs) and $j_\delta \sim \lambda\sigma_{\text{eq}}$ (for matter fluctuations). For $f\ll \sigma_{\text{eq}}$, the distribution $P(j_i)$ is therefore mostly determined by the matter fluctuations rather than by PBH torques. Only when $f \gtrsim \sigma_\text{eq}$ do the torque from other PBHs play a significant role. 

For PBH pairs with $j_i$ exceeding unity, the angular-momentum barrier prevents decoupling from the Hubble flow and binary formation already in RD. Again we note that the growth of angular momentum according to Eq.~\eqref{eq:jt} can also prevent binary formation in MD for lower initial $j_i < 1$, as we will be shown in the next Section.

\subsection{Properties of newly born PBH binaries}
\label{sec:newlybornPBHB}

\begin{figure}[h!]
\centering
\includegraphics[width=0.45\textwidth]{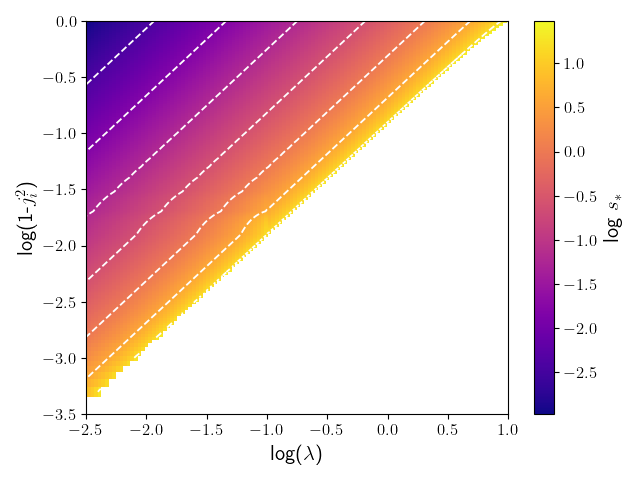}
\includegraphics[width=0.45\textwidth]{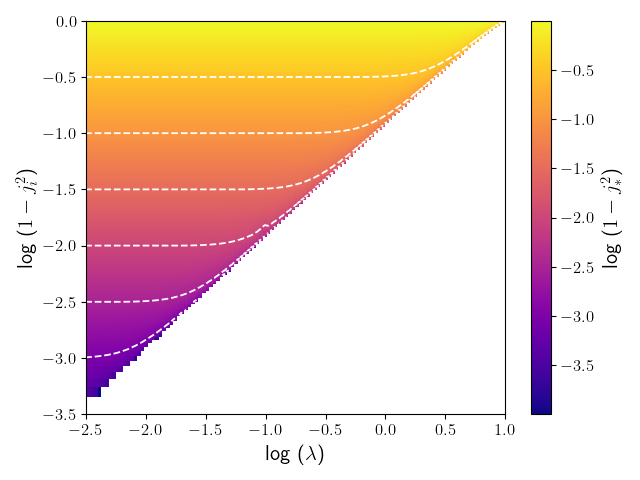}
\caption{Contour levels of $s_*$ (left panel) and $\log(1-j_*^2)$ (right panel) in the $\log\lambda$ - $\log(1-j_i^2)$ plane. A PBH pair with parameters in the white part of the parameter space does not decouple to form a binary. The left panel suggests the functional dependence $s_* = s_*(\frac{\lambda}{1-j_i^2})$, which turns out to hold only up to MR equality $\frac{\lambda}{1-j_i^2} \sim 1$. For $s_*>1$, the angular-momentum barrier grows as seen in the right panel. This growth hinders the formation of PBH binaries after $s_* \sim 30$.} 
\label{figure:sdec_data}
\end{figure}

We solve the equation of motion Eq.~(\ref{eq:EOM}) for various values of $\lambda$. The initial conditions always match the Hubble flow, that is, $X(\ms=0)=0$ and $X'(\ms=0)=1$. We take $\epsilon = 10^{-20}$ in order to accurately determine the scale factor $\ss(\lambda) = \lambda \ms_*(\lambda,j_i)$ at binary formation, which defines the formation redshift $z_*$ and formation time $t_*$. The corresponding initial semi-major axis and angular momentum of the PBH binary are $\as(\lambda) = \lambda x_i X_*(\lambda,j_i)/2$ and $j_* = j(s_*)$. Fig.~\ref{figure:sdec_data} shows the outcome of the numerical evaluation of Eq.~\eqref{eq:EOM} for $s_*$ and $j_*$ as a function of $\lambda$ and $j_i$.

The general trend seen in Fig.~\ref{figure:sdec_data} can be explained by the facts that (i) larger values of $\lambda$ imply larger initial separations $x_i$ and later decoupling, while (ii) larger values of $j_i$ increase the angular-momentum barrier and delay binary formation as well. From the left panel of Fig. \ref{figure:sdec_data}, one might conclude that the parameter $\frac{\lambda}{1-j_i^2}$ primarily determines PBH binary formation. However, this observation only holds until MR equality ($\frac{\lambda}{1-j_i^2} \sim 1$), after which the time-dependence of the angular momentum dominates in delaying (and ultimately in preventing) the decoupling of PBH pairs from the Hubble flow. After MR equality, the initial parameter $\frac{\lambda}{1-j_i^2}$ alone is therefore insufficient to predict the properties of the newly-born PBH binary. Furthermore, note the absence of PBH binary formation beyond $s_{*,max} \sim 30$. At this point, the reduced angular momentum of the system exceeds unity, i.e. $j(t) > 1$ even for $j_i = 0$, such that binary formation in this manner becomes impossible after this point in time. The critical value of $s_{*,max}\sim 30$ can be inferred as follows: the left panel of Fig. \ref{figure:sdec_data} shows that $\ss$ is maximum on a diagonal given by $\frac{\lambda}{1-j_i^2}\sim 10$. On this diagonal, setting $j_i = 0$ yields $\lambda \sim 10$ which, once substituted into Eq.~\eqref{eq:jt}, leads to $j(t)\simeq1$ when $s=s_{*,max} \sim 30$.

The approximate solutions $\ss(\lambda) \sim \lambda/3$ and $\as(\lambda) \sim 0.1\lambda x$ found by \cite{Ali_Ha_moud_2017} hold when binary formation occurs deep in the RD era, where the angular-momentum barrier is negligible. In this case, we have $\lambda, j_i \ll 1$ separately. These approximate solutions break down for binary formation around MR equality ($(\frac{\lambda}{1-j_i^2}) \sim 1$) or further in the MD era. The reduced angular momentum $j_*$ of the system at PBH binary formation follows from Eq.~\eqref{eq:jt}, i.e. $j_*=j_*(j_i,s_*(\lambda))$, while Eq.~\eqref{angularmomentumeccentricity} determines the initial eccentricity $e_*$ of the binary.

\begin{figure}[h!]
\centering
\includegraphics[width=1\textwidth]{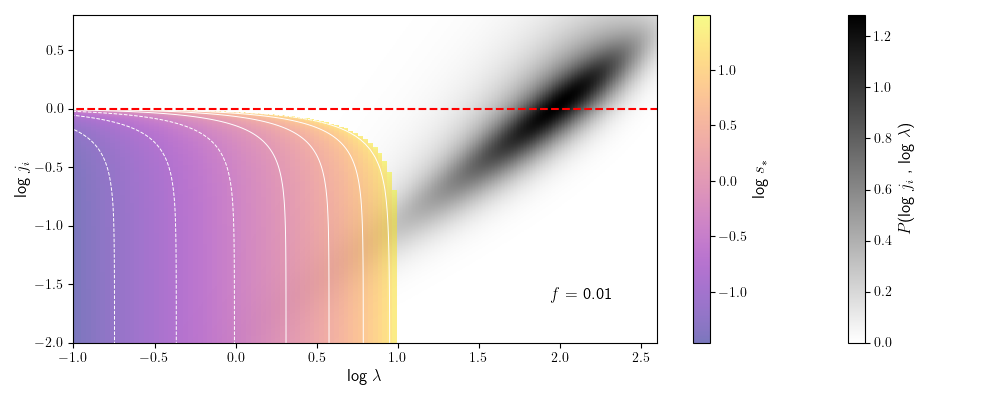}
\caption{The gray-shaded area shows the joint distribution $P(\log j_i , \log \lambda)$ given by Eq.~\eqref{eq:combi} for $f=0.01$. The red dashed line indicates where $j_i$ reaches unity. The angular-momentum barrier prevents PBH binary formation for large $\lambda\gtrsim 10$ even when $j_i<1$ due to the time dependence of $j(t)$. Contour levels of the scale factor $s_*$ are overlaid in color (as in Fig. \ref{figure:sdec_data}).}
\label{fig:secondinitial2dpdf}
\end{figure}

The properties of the newly-born PBH binaries are characterized by the distribution of the variables $(a_*,s_*,j_*)$, which are not independent. E.g. $a_*$ is completely determined once $j_*$ and $s_*$ are known. In particular, the joint distributions $P(j_*,\ss)$ and $P(j_*,\as)$ are fully specified by $P(j_i,\lambda)$, which is itself determined by $P(x_i)$ through
\begin{equation}
    P(j_i,\lambda) = P(j_i|\lambda)P(\lambda) \;.
    \label{eq:combi}
\end{equation}
The conditional PDF $P(j_i|\lambda)$ is given by Eq. \eqref{Pj} and
\begin{equation}
    P(\lambda) = P(x_i)\dv{x_i}{\lambda} = f e^{-f\lambda}\;.
\end{equation}
Note that this expression is independent of the PBH mass $m$. Combining this with the angular momentum distribution $P(j_i)$ discussed above, we construct the PDF $P(\log j_i , \log \lambda)$, as shown in Fig. \ref{fig:secondinitial2dpdf}.

As seen in Fig. \ref{fig:secondinitial2dpdf}, a significant fraction of $P(\log j_i, \log\lambda)$ extends beyond the part of parameter space that results in decoupling. These represent PBH pairs that will not form a binary, since for a given $j_i$, increasing $\lambda$ implies a stronger growth of the angular momentum barrier via Eq. \eqref{eq:jt}. The angular-momentum barrier is thus even more likely to prevent the PBH pair from forming a bound system. 

Transforming $(j_i,\lambda)$ to the variables $(s_*, a_*, j_*)$, we compute $P(j_*,a_*)$, and the cumulative distribution $P(<s_*)$ upon marginalizing $P(j_*,s_*)$ over $j_*$. Note that, whereas both $P(s_*,j_*)$ and $P(a_*,j_*)$ depend on $f$, only $P(a_*,j_*)$ depends also on the PBH mass $m$ through the relation $\as(\lambda) = \lambda x_i X_*(\lambda,j_i)/2$, with $x_i$ related to the random variable $\lambda$ according to Eq.~\eqref{lambdadef}.  

\begin{figure}[h!]
  \centering
    \includegraphics[width=0.6\textwidth]{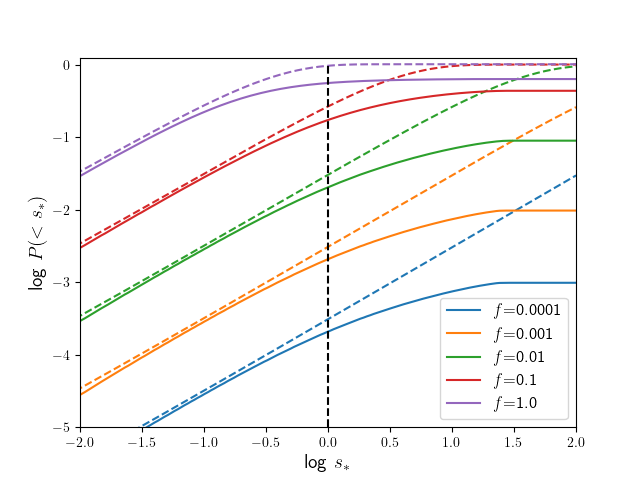}
  \caption{The cumulative distribution of the decoupling scale factor $P(<\ss)$. The dashed curves indicate the results assuming RD at all times and neglecting the angular-momentum barrier (a valid assumption in RD). The solid curves show the results for a universe with both matter and radiation, including the angular-momentum barrier. The vertical (black) dashed line marks MR equality. $P(<\ss)$ approaches a constant value $\eta_*(f)<1$ in the limit $s_* \rightarrow \infty$, which reflects the suppression of PBH binary formation due to the angular-momentum barrier for $\ss>s_{*,max} \simeq 30$. The asymptotic value of $P(< \ss)$ represents the total binary-formation probability $\eta_*$.}
  \label{fig:cdfs}
\end{figure}

\begin{figure}[h]
\centering
\includegraphics[width=1\textwidth]{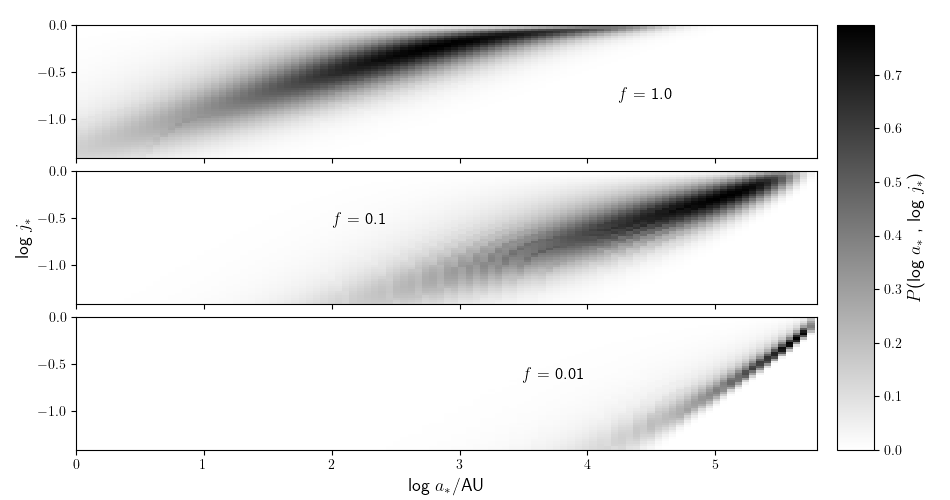}
\caption{$P(\text{log}a_*,\text{log}j_*)$ for $m=1$ and three choices for $f$. A different choice for $m$ only slightly shifts the PDF towards higher semi-major axes. For $f<0.01$, the distribution in $a_*,j_*$ highly resembles the $f=0.01$ (since the decoupling is mostly determined by the growing angular-momentum barrier early in the MD regime), with the difference mostly in $P(s_*)$ as can be seen from Fig. \ref{fig:cdfs}. All these PDFs have been normalized to 1 for visualization reasons.}
\label{fig:decouplingpdf}
\end{figure}

The cumulative distribution $P(<\ss)$ is shown as the solid curves in Fig.~\ref{fig:cdfs}, where only here and in Fig. \ref{fig:decouplingpdf} we include $f>0.01$ for illustrative purposes. The deviation from the RD expectations (represented by the dashed curves) emphasizes the importance of taking into the angular-momentum barrier for binary formation in the MD era, especially for PBH models with $f < 1$. The transition to MD suppresses binary formation in two ways. First of all via the growth of the angular-momentum barrier of Eq. \eqref{eq:jt}, and secondly due to the increase in the Universe's expansion rate in MD. Furthermore, we define $\eta_*$ as the fraction of PBHs that eventually decouple to form a binary. $\eta_*$ therefore equals the asymptotic value achieved by the cumulative PDF $P(<\ss)$. We observe that $\eta_*=\eta_*(f)$ is a decreasing function of $f$, with $\eta_*\sim 0.1$ for $f=0.01$. We will return to this in Section \S\ref{sec:PBHB}. 

Finally, Fig.~\ref{fig:decouplingpdf} shows the joint distribution $P(j_*,\as)$ for $m=1$ and three choices of~$f$. First, note that PBH binaries become wider as $f$ is lowered, since the initial separation $x_i$ increases when the PBH number density drops. Secondly, note that binary formation in MD is delayed due to the weakening of the effective gravitational attraction by the angular-momentum barrier (see Eq.~\eqref{eq:Newton}). For $f=1$, Fig.~\ref{fig:cdfs} shows that only a small fraction of PBH pairs decouples in MD. Hence, only a small part of the PBH binary population shows the effect of a delay in decoupling, visible as a tail towards wide circular orbits in the top panel of Fig.~\ref{fig:decouplingpdf}. For lower values of $f$, most PBH binaries form in the MD era, which implies that the delay caused by the angular-momentum barrier affects nearly all of the PBH binary population, shifting it to higher $a_*$. 


\section{Evolution of Galactic PBH binaries}
\label{sec:PBHevolution}

Having discussed the formation of PBH binary systems in the early Universe, we need to evolve the systems to the present epoch, where the PBH binaries trace the MW DM halo. In this evolution process, we deal with three possible processes: early disruption by matter nonlinearities (which turns out to be satisfied automatically), GW hardening, and stellar disruption in the MW halo. These processes will be discussed separately below. Furthermore, since PBH models in the (sub)solar mass range with fractions $f>0.01$ are severely constrained by microlensing events \cite{Mroz:2024mse}, we shall hereafter present results for $f\lesssim 0.01$ only.

\subsection{Early disruption by matter nonlinearities} 

For PBH binaries decoupling from the Hubble flow in MD, nonlinear structures on the scale of the binary semi-major axis can significantly alter the simple scenario considered here. Nonlinear structures may disrupt the formation of PBH binaries. Moreover, nonlinearities in the matter distribution will quickly trigger star formation and, thereby, the formation of compact stellar remnants which could mimic PBHs when $m$ is the solar mass range. To avoid these complications, we focus on "early-type" PBH binaries and discard those forming late in the MD era, when $\sigma_m$ computed at redshift $z_*$ and smoothed on scale $a_*$ exceeds unity.

We find that "early-type" PBH binaries forming at redshift $z_*\lesssim 20$ should be discarded. As can be seen in Fig. \ref{fig:cdfs}, the angular-momentum barrier prevents binary formation already for $z_*\lesssim 100$. This implies that nonlinearities in the high-redshift matter density field do not alter the PBH binary distribution under consideration here. 

\subsection{Hardening through GW emission}
\label{sec:gwhardening}

As stated previously, knowledge of the parameters $\lambda$ and $j_*$ is all that is needed to determine the properties of newly born PBH binaries characterized by the values of $(z_*,a_*,e_*)$. Conservation of energy and angular momentum implies that the semi-major axis $a$ and the eccentricity $e$ of the PBH binary do not evolve independently. Assuming that PBH binaries only evolve through (vacuum) GW emission in the quadrupole approximation, the joint evolution of $(a,e)$ follows from the coupled evolution equations for an eccentric orbit \cite{Peters:1963ux, Peters:1964qza, Maggiore:2007ulw}
\begin{equation}
    \begin{split}
        \dv{a}{t} &= -\frac{128}{5}\frac{G^3M^3}{c^5 a^3}\frac{1}{(1-e^2)^{7/2}}\left(1+\frac{73}{24}e^2+\frac{37}{96}e^4\right) \;,\\
        \dv{e}{t} &= -\frac{608}{15}\frac{G^3 M^3}{c^5 a^4}\frac{e}{(1-e^2)^{5/2}}\left(1+\frac{121}{304}e^2\right) \;. \\
    \end{split}
    \label{coupled_original}
\end{equation}
where $M = 2m\text{M}_{\odot}$ is the total mass of the binary system. Note that Eqs.~(\ref{coupled_original}) are orbit-averaged evolution equations. In our situation of highly separated, very eccentric orbits, the orbital phase of the system may impact the exact radiative power significantly. However given that we deal with a large population of binaries, we assume that population-wise it is sufficient to work with these orbit-averaged evolution equations nonetheless. From these equations, the coalescence time $t_{\text{coal}}$ of a binary with initial parameters $a_*, e_*$ is given by 
\begin{equation}
    t_{\text{coal}} = \frac{5}{256}\frac{c^5{a_*}^4}{G^3M^2\mu}F(e_*)
    \label{tcoal}
\end{equation}
where $\mu = \frac{1}{2}m\text{M}_{\odot}$ is the reduced mass. The auxiliary function $F(e_*)$ is defined in Appendix~\S\ref{sec:eccentricgwb}, where it is also shown that the system (\ref{coupled_original}) can be integrated to derive an analytical solution for the time needed to reach a new orbit with parameters $(a,e)$. This solution $t=t(a_*,e_*,e)$ is given by
\begin{equation}
    t(a_*,e_*,e) = \frac{15}{304}\frac{c^5{a_*}^4}{G^3\mu M^2}\frac{I(e_*)-I(e)}{{\mathcal{G}^4(e_*)}} \;.
    \label{integralsolve}
\end{equation}
The function $I(e)$ can be expressed in terms of Appell hypergeometric functions (see Appendix \ref{sec:appelsolution} for details). This expression can be (numerically) inverted to obtain the eccentricity $e(t,a_*,e_*)$ at time $t>t_*$ given the initial conditions $(a_*,e_*)$. The solution for $a=a(t,a_*,e_*)$ straightforwardly follows from Eq.~(\ref{eq:ae})


The system \eqref{coupled_original} involves two distinct timescales $t_a$ and $t_e$ characterizing the evolution of $a$ and $e$, respectively. $t_a$ depends on a higher power of $(1-e^2)$ and, therefore, is shorter than $t_e$. By neglecting $\mathcal{O}(1)$ contributions from the terms in the brackets in Eq.~\eqref{coupled_original}, the PBH binary evolves on a dynamical timescale $t_a$, which is initially
\begin{equation}
t_{a_*}\equiv \frac{c^5{a_*}^4}{G^3\mu M^2}\left(1-{e_*}^2\right)^{7/2}\;. 
\end{equation}
Only PBH binaries with initial conditions $(z_*,a_*,e_*)$ such that 
\begin{equation}
    t_{a_*} < \alpha t_H
\end{equation}
will significantly evolve under GW emission within the age of the Universe $t_0$. Here, $t_H$ is the present-day Hubble time and $\alpha\geq 1$ generically. For PBH binaries that have not merged by the present epoch, we choose to evolve in time only those systems with $\alpha=40$ in order to balance the computational cost (which becomes heavy when $\alpha$ is large) and the accuracy of the orbital evolution (which becomes poor when $\alpha$ is small). The PBH binaries with initial eccentricities and semi-major axes such that the binaries have already merged can be excluded from the present-day population.

\begin{figure}
\centering
\includegraphics[width=1\textwidth]{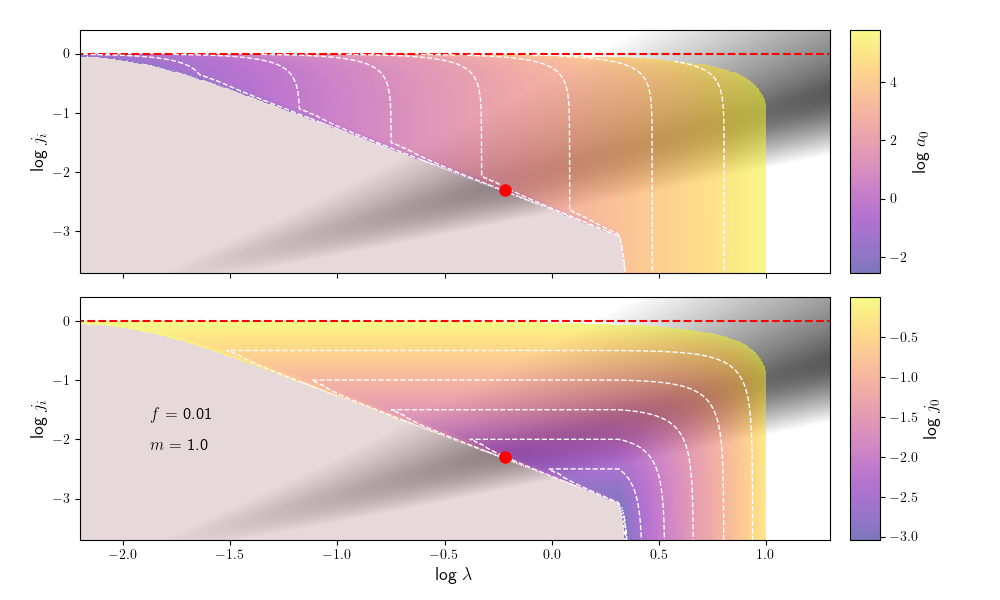}
\caption{Semi-major axis and reduced angular momentum after evolution through GW emission for $m=1, f=0.01$. Dashed (white) curves indicate contours of constant $a_0$ (top panel) and $j_0$ (bottom panel) in the $\lambda$ - $j_i$ plane. The pink-shaded area shows the initial parameter space leading to a merger by $t\leq t_0$. Mergers at $t=t_0$ have characteristic values of $j_i$ and $\lambda$ given approximately by the red dot. We have also overlaid the distribution $P( j_i,\lambda)$ as the grey-shaded area (as in Fig.~\ref{fig:secondinitial2dpdf}) to emphasize that only a small fraction of PBH binaries merge by the present epoch.}
\label{fig:evolution_nodisrupt}
\end{figure}

Fig.~\ref{fig:evolution_nodisrupt} displays the final (present-day) semi-major axis and reduced angular momentum obtained from the computation of the orbital evolution under GW emission solely, for a model with $m=1$ and $f=0.01$. The PBH binaries that merge at the present epoch are located near the red spot, on the diagonal separating the red-shaded area from the colored one.
The red spot approximately corresponds to the most probable values $a_* \sim 400 AU$ and $j_* \sim 0.003$ for a merger at $t=t_0$. This agrees with the predictions of \cite{Ali_Ha_moud_2017} (see their Fig.~4).
For log$\lambda \gtrsim 0.4$, evolution via GW emission is reduced because, in this part of the parameter space, the angular-momentum barrier increases so much that the characteristic GW evolution timescale becomes much larger than $t_H$.  

The overlap between $P( j_i,\lambda)$ (shown as the grey-shaded area) and the region of the initial parameter space leading to a merger by $t\leq t_0$ (shown as the pink-shaded area) emphasizes that only a small fraction of PBH binaries that ever formed have merged by today. Likewise, only a small fraction of the PBH binaries (those close to the region of merger) significantly evolve via GW emission. For $m=1$ and $f=0.01$, only about 4\% of the binaries have merged. This percentage does not change much for other choices of $m$ and $f$, the highest merger fraction being $\sim11$\% for $m=1000$ and $f=0.0001$. Consequently, we will disregard hierarchical mergers and hierarchical binary formation in this study. This is also in line with our assumption of a constant $f$.

\begin{figure}[h!]
\centering
\includegraphics[width=1\textwidth]{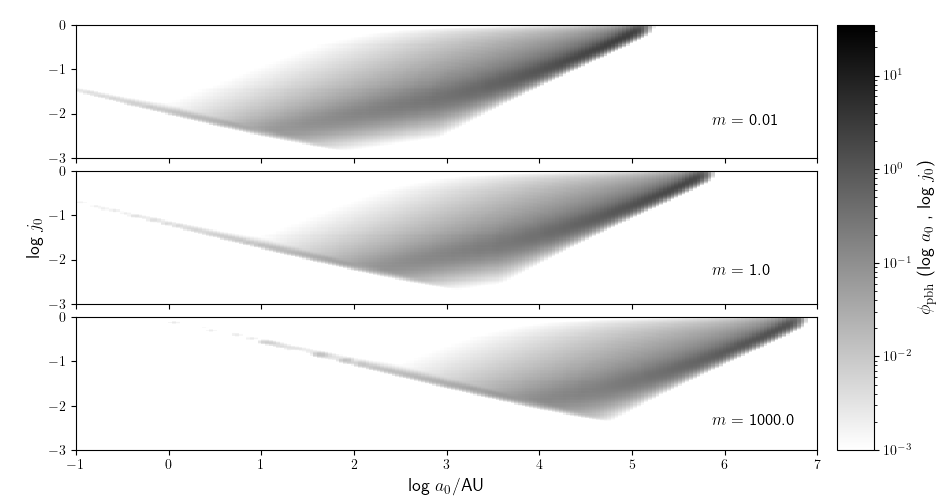}
\caption{The present-day distribution $\phi_\text{pbh}(a_0,j_0)$ (shown with a logarithmic color scheme) for PBH binaries assuming $f=0.01$ and three choices of $m$. The tail extending to smaller separations is caused by the loss of orbital energy through GW emission.}
\label{fig:finalpdf}
\end{figure}

At this point, we can evaluate the present-day PDF $\phi_\text{pbh}(a_0, j_0)$ of PBH binaries without the disruption effects that can take place in the MW halo (they will be discussed in Section \S\ref{sec:disruption}). In Fig. \ref{fig:finalpdf}, this PDF is shown for $f=0.01$ and three choices of $m$. As stated earlier, the distribution doesn't change noticeably for different choices of $f<0.01$, except for the number and fraction of PBH binaries.
The key point here is that present-day PBH binaries are typically very eccentric and very wide and, therefore, will emit GWs at very high harmonics and mostly at pericenter passage. Additionally, it is important to note that despite the high eccentricity of the binaries, their pericenter distance remains sufficiently large to prevent substantial relativistic corrections in their orbital evolution.

\subsection{Late-time disruption in the Milky Way halo}
\label{sec:disruption}

Once the MW halo has formed, the PBH binary can interact with MW stars and ``field'' PBHs (we ignore binary-binary interactions for simplicity), resulting in the disruption of the softest binaries. This effect depends on the stellar and PBH number density in the neighborhood of the PBH binary. As shown in Section \S\ref{sec:galdist}, field PBHs and MW stars are distributed in up to four components (bulge, halo, thin disk, and thick disk). Consequently, the present-day space distribution $\phi_\text{pbh}(\vr)$ of PBH binaries and the distribution $\phi(a_0,e_0)$ of orbital parameters are not independent anymore once binary disruption is taken into account. Furthermore, binary disruption is, in essence, probabilistic \cite[see][]{Goodman:1992cy,Ginat:2021ypu}. For simplicity, however, we shall ignore its probabilistic nature and treat instead PBH binary disruption in a pure deterministic manner. Finally, we have ignored additional sources of disruption \cite[reviewed in][]{Carr:2023tpt}\footnote{\label{fn:3body} We also do not consider the case where one of the binary members might be replaced by a third object during a binary-single encounter \cite[e.g.][]{Ginat:2021ypu}. For hard binaries, the binary-single scattering cross-section in the stellar halo is small anyway, but also for soft binaries, we don't take general hardening evolution processes by 3-body encounters into account \cite{Heggie:1992cz,GinatPerets2021b}. Next to this, as is noted in \cite{Raidal_2019}, for $f \sim 1$, nearly all initial binaries are expected to be disrupted by surrounding PBHs, based on simulations. Here $f \leq 0.01$, hence this is not the case here. Also \cite{Raidal_2019} includes bound systems of several PBHs that can form around MR equality when $f \sim 1$. In the case $f\ll 1$, bound systems of more than 2 PBHs are highly unlikely to form.}. In particular, we have neglected dynamical friction (DF) produced by the PDM component on the evolution of PBH binaries in the MW halo since, in the regime where DF is relevant, characteristic timescales for orbital-energy dissipation are much longer than disruption timescales (see Appendix \S\ref{sec:DF}). We have also ignored Galactic tides \cite{HeislerTremaine1986, GrishinPerets2022, ModakHamilton2023} because the time-scale of this process for $m\approx 1$ is too long for all but the widest binaries in the Galactic halo \cite{GrishinPerets2022}.  

To encode PBH binary disruption, we introduce a function $\Pi_\text{d} = \Pi_\text{d}(\vr,a_0)$, which equals $0$ if a PBH binary with semi-major axis $a_0$ is disrupted at position $\vr$, and $1$ otherwise. $\Pi_\text{d}$ depends also on the local stellar density $\rho_*(\vr)$ (calculated from the stellar MW profile described in Section~\S\ref{sec:galdist}) and PBH density $\rho_\text{pbh}(\vr)$ (caclulated from the NFW profile Eq.~(\ref{haloprof})). We assume a total stellar mass in the MW of $5 \times 10^{10}$ M$_\odot$ \cite{Flynn:2006tm,Bovy:2013raa,Licquia2015,Cautun2020}, which (only for the disruption procedure) is taken to be fully made up by potentially-disrupting stars.

To define $\Pi_\text{d}$, we divide the present-day PBH binaries into two categories, based on their binding energy $E_b = Gm^2\text{M}_{\odot}^2/2a_0$ relative to the velocity dispersion of PBHs and MW stars, which we take to be $\sigma_v = 100$ km/s regardless of $\vr$. For simplicity, we also assume encounters with MW stars of mass $m_* = 1 $M$_{\odot}$ solely. Hard binaries, for which $E_b > m_*\sigma_v^2$, are not affected by any disruption effect \cite{Ginat:2021ypu}.
However, soft binaries for which $E_b \leq m_*\sigma_v^2$ may be disrupted via two distinct channels (following Chapter 7 of \cite{2008gady.book.....B}):
\begin{itemize}
    \item Ionization, with a rate $R_I  \simeq 27.03\,\frac{G\rho_*a_0}{\sigma_v}$
    \item Evaporation, with a rate $R_E \simeq 16.39\,\frac{\ln(\Lambda)G\rho_*a_0}{\sigma_v}$
\end{itemize}
Here, $\ln(\Lambda)$ is a Coulomb logarithm with $\Lambda \sim 0.6\times\frac{m_*\sigma_v^2}{E_b}$. Hence, evaporation dominates for very soft binaries, while both processes contribute more or less equally when the binary is harder. A soft binary is assumed to be disrupted when $R_I+R_E > 1/t_H$. Note that this prescription is independent of the orbital phase and eccentricity.\footnote{The standard calculation of disruption rates \cite{binney_galactic_1987} assumes that the impact parameter between a star and a PBH is much smaller than the binary separation, which may break down at pericenter due to the high eccentricities. However, since highly eccentric PBH binaries spend only a brief fraction of their orbit near pericenter, this approximation remains valid for most of their evolution. This is further supported by the binary-single scattering experiments of \cite[][eq. (5.8)]{HutBahcall1983}, valid for arbitrary eccentricity, showing that the cross-section for disruption does not change considerably. Ultimately, our orbit-averaged framework is justified by the large population of binaries.}


\begin{figure}
\centering
\includegraphics[width=1\textwidth]{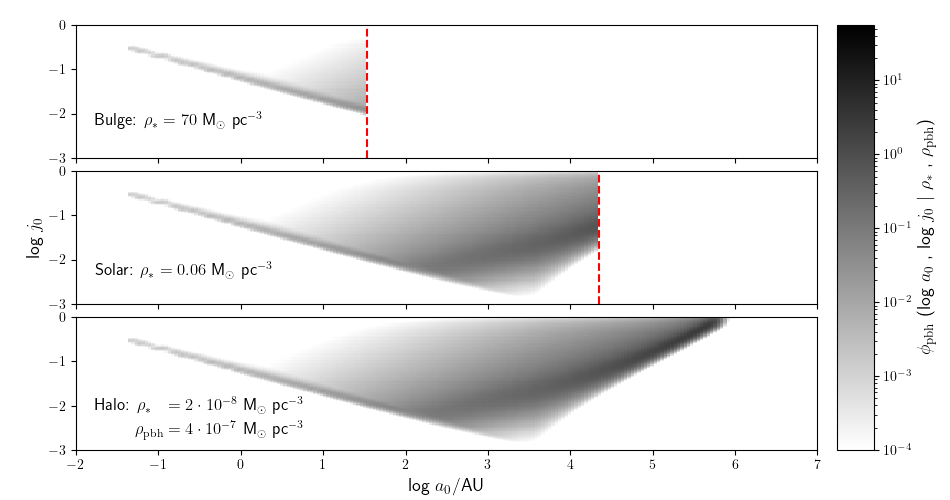}
\caption{The effect of disruption on the parameter distribution $\phi(a_0,j_0)$ at three locations in the MW: the bulge, disk, and halo for a model with $m=1, f= 0.01$. The region to the right of the red dashed line is disrupted at the respective locations. Note the logarithmic color scheme, showing that nearly all PBH binaries are disrupted in the bulge.}
  \label{fig:evolution_disrupt}
\end{figure}

For illustration, let us quantify how the disruption process affects the distribution shown in Fig.~\ref{fig:finalpdf}, which includes only binary hardening through GW emission. Fig.~\ref{fig:evolution_disrupt} shows the impact of disruption in three different MW environments or, equivalently, stellar densities: $\rho_* = 70 \text{M}_{\odot} \text{pc}^{-3}$ (for the MW bulge); $\rho_* = 0.06 \text{M}_{\odot}\text{pc}^{-3}$ (appropriate to the disk in the solar neighborhood); and $\rho_* = 2\times 10^{-8} \text{M}_{\odot}  \text{pc}^{-3}$ (for the stellar halo). In the latter case, we also take into account a contribution from field PBHs with density $\rho_\text{pbh}=4\times 10^{-7} \text{M}_{\odot}  \text{pc}^{-3}$. In the halo, both the stellar and field PBH densities are too low to cause significant disruption effects\footnote{In \cite{Young:2020scc}, it has already been shown that fly-by events of single PBHs are expected only to have a subdominant effect on PBH binary properties.}.The resulting distribution $\phi_\text{disrupt}(a_0,j_0)$ shown in Fig.~\ref{fig:evolution_disrupt} is identical to the distribution before including disruption except for a cut in semi-major axis. Results are shown for $f=0.01$ and  $m=1$.

\section{Galactic distributions}
\label{sec:galdist}

In this Section, we compute the spatial distribution of Galactic PBH binaries. Since the disruption of PBH binaries depends on the local stellar density, we begin with a model for the spatial distribution of MW stars. This profile serves a dual purpose, as it is also essential in the calculation of the GW foreground produced by Galactic DWDs. 

\subsection{Galactic Double White Dwarfs}
\label{sec:DWD}

The dominant component of the astrophysical compact binaries in the MW are $N_\text{dwds}\simeq 10^8$ DWDs \citep[e.g.][]{Korol:2021pun}. They provide a foreground to the GW signal produced by PBH binaries in the millihertz (mHz) frequency range probed by the LISA experiment. Hence accurate modeling of the DWD population is essential.  

Assuming that the DWD population in the MW traces the stellar component, we follow \cite{Cooper:2009kx,Lin:2019yux,Valenti_2016,AllendePrieto:2009vht} and express the probability $\phi_\text{dwd}(\vr_g)$ for a DWD to be located at a separation $\vr_g$ away from the MW Galactic Center as
\begin{equation}
    \phi_\text{dwd}(\vr_g) = f_\text{td}\, \phi_\text{td}(\vr_g) + f_\text{TD}\, \phi_\text{TD}(\vr_g) + f_\text{b}\, \phi_\text{b}(\vr_g) + f_{\text{h}_*}\, \phi_{\text{h}_*}(\vr_g) \;,
\end{equation}
where the probability densities $\phi_\text{td}(\vr_g)$, $\phi_\text{TD}(\vr_g)$, $\phi_\text{b}(\vr_g)$ and  $\phi_{\text{h}_*}(\vr_g)$ are constructed from the Galactic profiles of the thin disk, thick disk, bulge and stellar halo, respectively. Furthermore, $f_i$ are the relative weight of each component. 

The disks decay exponentially in the Galactic plane $x$ -- $y$, while they quickly fall off along the vertical $z-$direction, where $(x,y,z)$ are Cartesian coordinates relative to the Galactic Center:
\begin{align}
    \phi_\text{td}(\vr_g) &\propto \frac{e^{-\sqrt{x^2+y^2}/H_\text{d}}}{\cosh(z/h_\text{td})}
    \label{tdprof} \\
    \phi_\text{TD}(\vr_g) &\propto \frac{e^{-\sqrt{x^2+y^2}/H_\text{d}}}{\cosh(z/h_\text{TD})}
    \label{TDprof} \;.
\end{align}
Their densities are expressed as a function of $x^2+y^2$ because of cylindrical symmetry. The central bulge, on the contrary, is modeled as a simple radial exponential profile,
\begin{equation}
    \phi_\text{b}(\vr_g) \propto e^{-r_g/H_\text{b}} 
    \label{bulgeprof}
\end{equation}
where $r_g=|\vr_g|$ is the radial distance from the Galactic Center. Finally, a small fraction of DWDs lives in the MW stellar halo, modeled as a Navarro-Frenk-White (NFW) profile \cite{Lin:2019yux}:
\begin{equation}
    \phi_{\text{h}_*}(\vr_g)\propto \frac{R_{\text{h}_*}}{r_g}\left(1 + \frac{r_g}{R_{\text{h}_*}}\right)^{-2}
    \label{stellarhaloprof}
\end{equation}
The relative weights $f_i$ of each component and the parameter values are summarized in Table~\ref{parametertab}. We assume there are no stars beyond a separation of $100\kpc$ from the Galactic Center. 

\begin{table}
    \centering
    \caption{Parameters of the Galactic model spelled out in Eqs.~(\ref{tdprof}-\ref{stellarhaloprof}), based on \cite{Cooper:2009kx,Lin:2019yux,Valenti_2016,AllendePrieto:2009vht}.}
    \begin{tabular}{|l|l||l|l|}
        \multicolumn{2}{c}{Length Scales (kpc)} & \multicolumn{2}{c}{Relative Weights}\\ \hline
        $h_\text{td}$ & 0.3 & $f_\text{td}$ & 0.84 \\ \hline
        $h_\text{TD}$ & 1 & $f_\text{TD}$ & 0.05 \\ \hline
        $H_\text{d}$ & 3 & ~ &  ~ \\ \hline
        $H_\text{b}$ & 0.15 & $f_\text{b}$ & 0.1 \\ \hline
        $R_{\text{h}_*}$ & 8 & $f_{\text{h}_*}$ & 0.01 \\ \hline
    \end{tabular}
    \label{parametertab}
\end{table}

\begin{figure}[h!]
    \includegraphics[width=0.5\textwidth]{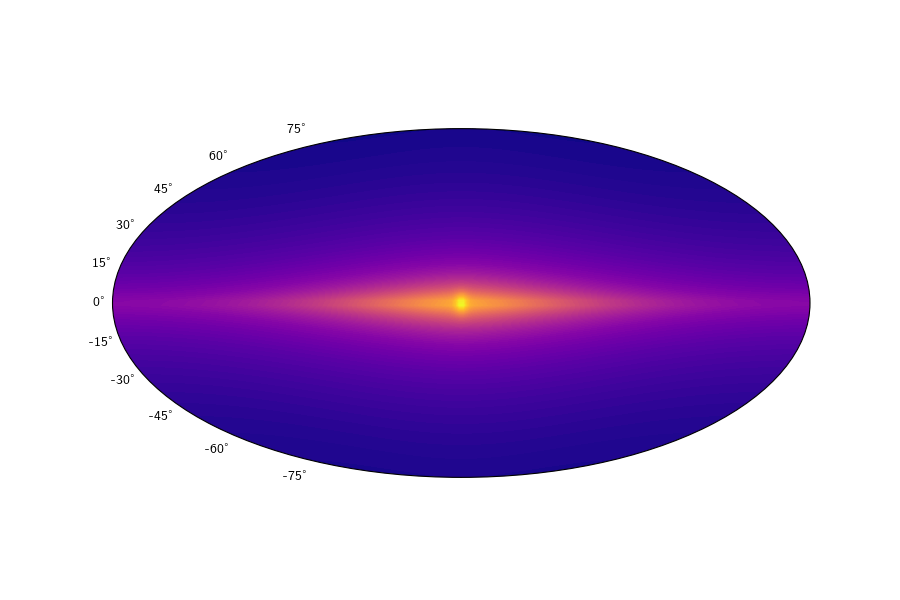}   
    \includegraphics[width=0.5\textwidth]{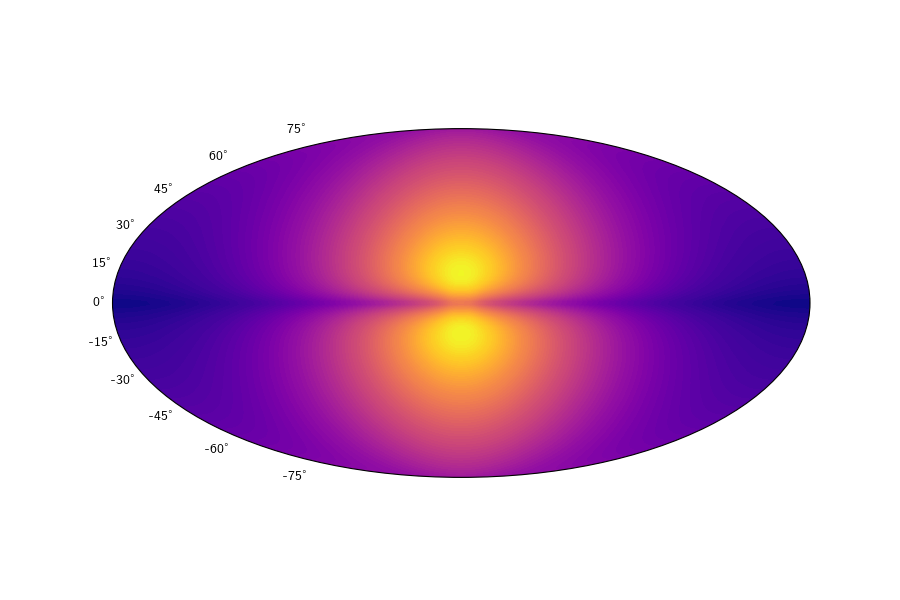}
  \caption{Angular maps of the sky density probability (i.e. integrated along the line of sight) of Galactic DWDs (left panel) and PBH binaries for $m=1$, $f=0.01$ (right panel). We observe that, due to stellar disruption, PBH binaries are mostly located in the halo, while DWDs are mostly concentrated in the disk and central bulge. }
  \label{fig:DWDPBHmap}
\end{figure}

The spatial distribution of Galactic DWDs is shown in the left panel of Fig.~\ref{fig:DWDPBHmap} in Galactic Coordinates (i.e. as viewed from the Sun). Note that the position of the Sun in the Cartesian coordinates used above is $\vr_{g,\odot}=(8249,0,20.8)$ pc \cite{Bennett2019, GravityCollab21}, with the $x-$axis pointing towards the solar system.

To model the GW power emitted by Galactic DWDs, we will also need their present-day semi-major axis and mass distributions. In this regard, we assume that the two white dwarf masses $\mathfrak{m}_1$ and $\mathfrak{m}_2$ follow a Gaussian distribution with mean $0.6$M$_{\odot}$ and standard deviation $0.1$M$_{\odot}$ \cite{Maoz:2016bxg}. Furthermore, we assume circular orbits with a semi-major axis $a_\text{dwd}$ distributed according to the broken power law model of \cite{Korol:2021pun} \footnote{If one would evolve the DWD population in time (as is the approach of \cite{Ginat:2019aed,Ginat:2023eto}), one could start from the initial binary separation distribution proposed by \cite{Maoz:2016bxg}, which is the power law $\phi(a_\text{dwd})\propto a^{-1.3}$.}:
\begin{equation}
    \phi(a_\text{dwd}) \propto x^{4+\beta}\bigg[\Big(1+x^{-4}\Big)^{\frac{\beta+1}{4}}-1\bigg]
\end{equation}
with $x \equiv a_\text{dwd}/0.01 \text{AU}$ and $\beta = -1.3$ and the distribution is ranging from a minimum separation of $14\cdot 10^3$ km up to a maximum separation of $0.05$ AU. Note that the DWDs are therefore much harder than the PBH binaries, and we do not have to take disruption effects into account. As is also the case for the PBH binaries, the orbital inclination $\cos \imath$ of the DWDs is uniformly distributed. 

\subsection{Galactic PBH binaries}
\label{sec:smoothDM}

For the PBH fractions $f\lesssim 0.01$ considered here, the PDM component dominates the MW halo. We assume that the PDM distribution of the MW follows an NFW profile with a characteristic radius $R_\text{h,dm} = 20\kpc$, such that the total MW mass is $2\times10^{12}$ M$_{\odot}$ and the local PBH+PDM density in the solar neighborhood is $\sim 0.01$ M$_{\odot}$pc$^{-3}$, i.e.
\begin{equation}
    \phi_\text{h,dm}(\vr_g)\propto \frac{R_\text{h,dm}}{r_g}\left(1 + \frac{r_g}{R_\text{h,dm}}\right)^{-2}
    \label{haloprof}
\end{equation}
where $r = |\vr|$. Since PBHs are initially a Poisson sampling of the adiabatic mode, they behave as test particles advected by the PDM component. Therefore, we shall assume that the gravitational collapse and virialization of the MW halo lead to a distribution of PBHs that approximately traces the MW halo of particle DM (a small fraction of them could end up in globular clusters, see \cite{Vanzan:2024wwc}). Hence we take the PBH spatial distribution to extend to the virial radius $R_{\rm vir} \simeq 200\kpc$. However, the position-dependent disruption of PBH binaries by stars and field PBHs in the MW will cause their present-day distribution to deviate from the NFW profile. Therefore, the present-day spatial profile of PBH binaries in the MW is given by
\begin{equation}
    \phi_\text{pbh}(\vr_g) = \frac{1}{\eta_\text{d}}\int\dd a_0 \; \phi_\text{h,dm}(\vr_g)\, \phi(a_0)\,\Pi_\text{d}(a_0,\vr_g)
\end{equation}
where the distribution function $\Pi_\text{d}(a_0,\vr_g)$ is defined in Section \ref{sec:disruption}, the normalization factor $\eta_\text{d}$ is defined below in Eq. \eqref{nondis} and $\phi(a_0)$ is computed by marginalizing $\phi(a_0,j_0)$ over $j_0$, which can be done since disruption is taken to be independent of the eccentricity of the binary. The resulting spatial distribution of PBH binaries is shown in the right panel of Fig.~\ref{fig:DWDPBHmap} for an observer at the Sun. The symmetry of the PBH binary distribution around the vertical axis reflects the axial symmetry of the stellar components responsible for the disruption of PBH binaries in the Galactic disk. 

\subsection{Number of Galactic PBH binaries}
\label{sec:PBHB}

The number $N_0$ of PBH binaries currently present in the MW halo depends on $m$ and $f$ according to Eq. \eqref{numberpbhbs}. Three distinct physical effects determine the value of $\eta_0$:
\begin{itemize}
\item Only a fraction $\eta_*(f)$ of all the PBHs form binaries (see~\S\ref{sec:PBHformation}). This is exemplified in Fig.~\ref{fig:cdfs}. For low values of $f$, the large distance between nearest neighbors and the angular-momentum barrier counteracts binary formation.
\item PBH binaries may merge due to the dissipation of orbital energy by GW emission (see~\S\ref{sec:gwhardening}). $\eta_\text{merge}$ will denote the fraction of PBH binaries that have not merged by the present epoch.
\item Galactic PBH binaries can be disrupted via stellar encounters  (see~\S\ref{sec:disruption}). The fraction $\eta_\text{d}$ of surviving binaries integrated over the MW halo can be expressed as
\begin{equation}
   \eta_\text{d} = \int\! \dd a_0\, \dd \vr_g\,\phi(a_0)\,\phi_{\text{h,dm}}(\vr_g)\,\Pi_\text{d}(a_0,\vr_g)\;.
  \label{nondis}
\end{equation}
\end{itemize}

\begin{figure}[h!]
    \includegraphics[width=1\textwidth]{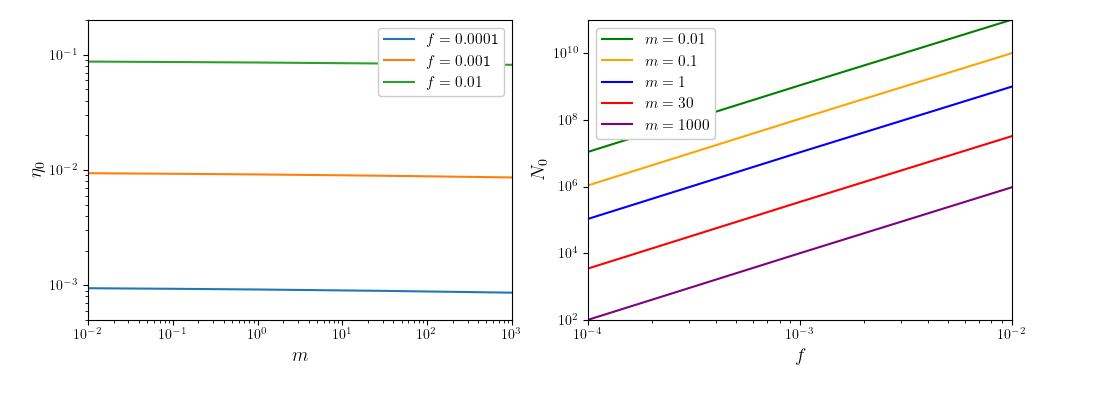}   
  \caption{\textit{Left panel:} Fraction $\eta_0$ of PBHs in the MW that live in a binary system. $\eta_0$ is mainly determined by $\eta_* = \eta_*(f)$, with a weak $m$-dependence arising through $\eta_\text{merge}$ and $\eta_\text{d}$ \textit{Right panel:} total number $N_0$ of Galactic PBH binaries for different choices of $m$ and $f$.}
  \label{fig:PBHnumber}
\end{figure}

Combining all these effects, the present-day fraction $\eta_0$ of MW PBHs in binary systems, relative to the total number of PBHs (see Eq.~\ref{numberpbhbs}) reads
\begin{equation}
\eta_0 = \eta_* \cdot \eta_\text{merge} \cdot \eta_\text{d}
\label{eq:toteta}
\end{equation}
Since the dependence of $\eta_0$ on $m$ and $f$ is known, the degeneracy in Eq.~\eqref{numberpbhbs} between $\eta_0$ and $f$ is now broken. The fractions $\eta_\text{merge}$ and $\eta_\text{d}$ are larger than $\sim 0.9$ and $\sim 0.98$ respectively for the values of $m$ and $f$ under consideration. This implies that $\eta_0$ is primarily determined by $\eta_*=\eta_*(f)$, which is a function of $f$ solely. The left panel of Fig.~\ref{fig:PBHnumber} exemplifies the weak dependence of $\eta_0$ on $m$. The right panel of Fig.~\ref{fig:PBHnumber} demonstrates that there are still regions of the $m$ - $f$ parameter space for which $N_0$ is comparable to or larger than the number $N_\text{dwd}\sim 10^8$ of Galactic DWDs. 

To conclude, note that we have assumed throughout that both $m$ and $f$ remain constant in time. In particular, we have neglected the growth of PBH mass via merger and accretion and the possibility that second-generation PBH binaries may form. These evolution effects will effectively lower $f$ in time. However, since we have found that $1-\eta_\text{merge}$ (which can be thought of as a proxy for the PBH merger rate) to be quite low, these effects will likely result in small corrections to the simplified scenario considered here. 

\newpage
\section{Detecting Galactic PBH binaries}
\label{sec:GWB}

In this Section, we explore the feasibility of detecting a GW background arising from the hypothetical population of Galactic PBH binaries, focusing on the LISA experiment. The detectability of this background depends on the level of the confusion noise of unresolved Galactic DWDs. We will not consider e.g. the GW signal produced by extragalactic DWDs \cite{Staelens:2023xjn}.

\subsection{GW Energy Density}

We compute the GW energy density (per logarithmic frequency interval $\dd \ln\nu$)
\begin{equation}
    \Omega_\text{gw}(\nu) \equiv \frac{1}{\rho_c}\dv{\rho_\text{gw}}{\ln \nu}
\end{equation}
sourced by Galactic DWDs and PBH binaries based on their present-day parameter distributions (unlike, e.g. \cite{Ginat:2023eto}, who expressed $\Omega_\text{gw}$ in terms of the source parameter distribution at formation time). 
Here, $\nu$ is the observed frequency of the gravitational waves. 
We gather the source parameters in the vector $\vxi$. Accounting for the sky-dependence of the signal and the orbital inclination $\imath$ of the binary sources, $\Omega_\text{gw}(\nu)$ is fully specified by the probability distribution $\phi(\vr,\vxi,\imath)$, where $\vr$ is the separation vector between the Earth and the source. For a uniformly distributed $\cos\imath$, we can write
\begin{equation}
\phi(\vr,\vxi,\imath) \dd\vr\,\dd\vxi\,\dd\cos\imath= \frac{1}{2}\phi(\vr,\vxi)\dd\vr\,\dd\vxi\,\dd\cos\imath \;.
\end{equation}
With these definitions, $\Omega_\text{gw}(\nu)$ can be expressed as \cite[e.g.][]{phinney:2001,Ginat:2023eto}
\begin{equation}
\begin{split}
     \Omega_\text{gw}(\nu)  = \frac{8\pi^2N\nu^3}{3{H_0}^2 F}\expval{\abs{\tilde{h}_{\nu,\imath}}^2}_{\vr,\vxi,\imath} = \frac{8\pi^2N\nu^3}{3{H_0}^2 F} \int \dd\vr\,\dd\vxi \,\phi(\vr,\vxi)\,\expval{\abs{\tilde{h}_{\nu,\imath}}^2}_\imath \;.
\label{omega_start}
\end{split}
\end{equation}
Here, $N$ is the total number of sources, and $\tilde{h}_{\nu,\imath}\equiv\tilde{h}_{\nu}(\vr,\vxi,\imath) $ is the discrete Fourier transform of the GW signal produced by a single source in the MW. We keep the dependence on $\imath$ explicit in the shorthand notation for reasons that will become clear below. The notation $\big\langle \dots\big\rangle_X$ indicates an ensemble average over the parameter $X$. Furthermore, the division by the angular form factor $F \equiv \expval{F^+}_{\hat{\boldsymbol{n}}} +\expval{F^\times}_{\hat{\boldsymbol{n}}}$ of the GW detector cancels out a similar factor in $\langle\lvert\tilde{h}_{\nu,\imath}\lvert^2\rangle_\imath$ and ensures that $\Omega_\text{gw}(\nu)$ is independent of the detector configuration.

For the LISA experiment, we have 
\begin{equation}
    \mathcal{R} \equiv \expval{F^+}_{\hat{\boldsymbol{n}}} = \expval{F^\times}_{\hat{\boldsymbol{n}}} \simeq \frac{3}{10} \;,
    \label{rfac}
\end{equation}
where $\mathcal{R}$ is the signal response function averaged over the sky and the polarization of the GWs. At higher order, $\mathcal{R}$ is frequency dependent. We shall stick to the first order approximation, valid up to frequencies $\lesssim 10^{-2}$ Hz above which the sensitivity of the LISA interferometer is significantly reduced \cite{Robson:2018ifk}. Therefore, Eq.~\eqref{rfac} implies $F = 2\mathcal{R} = 3/5$ for LISA (note that $F=2/5$ for the LIGO experiment). 

When the source emits at a single frequency as is the case for circular binaries, the discrete Fourier transforms $\tilde{h}_{\nu,\imath}$ are related to the Fourier modes $\tilde{h}(\nu_2,\imath)\equiv \tilde{h}(\nu_2,r,\vxi,\imath)$ of the whole continuous signal of a single source through \citep[e.g.][]{Ginat:2023eto}
\begin{equation}
    \tilde{h}_{\nu,\imath} = \tilde{h}(\nu_2,\imath)\, \frac{\Pi_T(\nu)}{\sqrt{T}}
\end{equation}
where $\nu$ is the observed frequency, $\nu_2$ is the (quadrupole) GW frequency of the source, equaling $\nu_2\equiv 2\nu_0(\vxi)$ with $\nu_0$ the orbital frequency. Furthermore, $\Pi_T(\nu)$ is the window function of an experiment with duration time $T$. On taking $T \rightarrow 0$, we have
\begin{equation}
   \lim_{T\rightarrow 0} \frac{\Pi_T(\nu)}{T} = \abs{\dv{\nu}{t}}\,\delta\big(\nu-\nu_2\big) \;.
   \label{limitwindow}
\end{equation}
This is the relevant limit in our case too, where the sources do not evolve significantly throughout the observational run. The Fourier modes are then given by
\begin{equation}
    \tilde{h}(\nu_2,\imath) = h_0(\nu_2,r,\vxi)\,e^{\text{i}\Psi(\vxi)}\,\mathcal{F}(\imath)\;,
\end{equation}
where $h_0(\nu_2,r,\vxi)$ is the amplitude of the inspiral waveform (to be specified shortly) and $e^{\text{i}\Psi(\vxi)}$ is its phase. The latter cancels out in the expression of $\Omega_\text{gw}(\nu)$. Moreover, the complex form factor $\mathcal{F}(\imath)$ depends on the orbital inclination. It can be approximated by~\cite{Robson:2018ifk}~\footnote{In this way $\tilde{h}(\nu_2,\imath)$ is independent of the angular position on the sky. Especially in the case of anisotropic sources, the total strain is correlated with the angular sky location, depending on the detector orientation. A more precise method would be to use the complicated separate polarization form factors $F^+, F^\times$, as outlined in e.g. \cite{Cornish:2001bb}. In this work, we neglect this technicality for simplicity. This is also justified by the added uncertainty around LISA's orientation and finalized design.}:
\begin{equation}
    \mathcal{F}(\imath) = \sqrt{\mathcal{R}}\left(\frac{1+\cos^2\imath}{2} + \text{i}\cos \imath\right)
    \label{approxinc}
\end{equation}
Averaging over the inclination angle $0\leq \imath<\pi$ gives
\begin{equation}
    \expval{\mathcal{F}(\imath)\mathcal{F}^*(\imath)}_\imath = \frac{4}{5} \mathcal{R} = \frac{2}{5} F \;.
\end{equation}
Combining all the above expressions, we arrive at
\begin{equation}
\begin{split}
     \Omega_\text{gw}(\nu) = \frac{16\pi^2N\nu^3}{15{H_0}^2} \int\! \dd\vr\,\dd\vxi \, \phi(\vr,\vxi)\,\abs{\dv{\nu}{t}}\,\delta(\nu-\nu_2)\,{h_0}^2(\nu_2,r,\vxi)
\label{omega_def}
\end{split}
\end{equation}
The dependence on the sky direction arises only through $\phi(\vr,\vxi)$, which encodes the spatial density of sources in the MW. One should bear in mind that Eq.~\eqref{omega_def} is strictly speaking only valid for circular orbits. We will generalize this expression to eccentric orbits in Section \S\ref{sec:PBHresults} when we discuss PBH binaries.

\subsection{Confusion noise from unresolved Galactic DWDs}
\label{sec:DWDnoise}

The confusion noise from unresolved Galactic DWDs is computed by assuming circular orbits, with the waveform given by
\begin{equation}
\label{circwave}
    h_0(\nu_2,r,\vxi) = \frac{1}{\pi^{2/3}}\sqrt{\frac{5}{24}}\frac{c}{r}\left(\frac{GM_c}{c^3}\right)^{5/6}\nu_2^{-7/6}
\end{equation}
where $M_c=(m_1m_2)^{3/5}(m_1+m_2)^{-1/5}$ is the chirp mass of the binary system. The parameter vector $\vxi_\text{dwd}$ of the DWD is composed of the semi-major axis $a_\text{dwd}$, the masses $\mathfrak{m}_1$ and $\mathfrak{m}_2$ of the individual white dwarfs, and the spatial location $\vr$ of the system in Galactic Coordinates. The distributions for these parameters are given in Section \ref{sec:DWD}.

We use $a_\text{dwd}$ to eliminate the Dirac delta in Eq.~\eqref{omega_def}. As a result, the GW energy density $\Omega_\text{dwd}(\nu)$ of the Galactic DWD foreground reads
\begin{align}
    \label{omega_dwd}
     \Omega_\text{dwd}(\nu) &= \frac{16\pi^2N_\text{dwd}\nu^3}{15{H_0}^2} \int \dd\vr\,\dd\vxi_\text{dwd} \,\phi_\text{dwd}(\vr,\vxi_\text{dwd})\,\abs{\dv{\nu}{t}}\,\delta(\nu-\nu_2)\,{h_0}^2(\nu_2,r,\vxi_\text{dwd})
     \\
     &= \frac{16\pi^2N_\text{dwd}\nu^3}{15{H_0}^2} \int \dd\vr\,\dd \mathfrak{m}_1\dd \mathfrak{m}_2 \,\phi_\text{dwd}(\vr,\vxi_\text{dwd})   \,\abs{\dv{a_\text{dwd}}{t}}{h_0}^2(\nu,r,\vxi_\text{dwd})\Bigg|_{a_\text{dwd}=a_\text{dwd}(\nu,\mathfrak{m}_1,\mathfrak{m}_2)} \nonumber \;,
\end{align}
where $\nu_2\equiv \nu_2(\vxi_\text{dwd})$, $a_\text{dwd}$ is fixed by the other parameters. Kepler's law and the relation $\nu = 2\nu_o$ between $\nu$ and the orbital frequency $\nu_o$ appropriate for circular orbits give 
\begin{equation}
    a_\text{dwd}(\nu,\mathfrak{m}_1,\mathfrak{m}_2) = \left(\frac{G(\mathfrak{m}_1+\mathfrak{m}_2)}{\pi^2}\right)^{1/3}\nu^{-2/3} \;.
\end{equation}
Combining this expression with orbital decay timescale via GW emission yields
\begin{equation}
    \abs{\dv{a_\text{dwd}}{t}} = \frac{64\pi^2}{5} \frac{G^2\mathfrak{m}_1\mathfrak{m}_2}{c^5} \nu^2 \;.
\end{equation}
Some of the Galactic DWDs will significantly exceed the LISA noise level. Their signal will be subtracted from the total GW strain to produce a confusion noise of unresolved DWDs with a GW energy density $\Omega_\text{ndwd}(\nu)\leq \Omega_\text{dwd}(\nu)$.

To implement the extraction of loud DWDs, and thereby maximizing the SNR of the PBH binary GW signal, let us first specify the LISA sensitivity, which is encoded in the effective noise power spectral density $S_\text{n}(\nu)$ given by
\begin{equation}
    S_\text{n}(\nu) \equiv \frac{P_n(\nu)}{\mathcal{R}} \;.
\end{equation}
Here, $P_n(\nu)$ the power spectral density of the detector noise. On substituting $\mathcal{R}= 3/10$, we obtain the following analytic fit for the noise \cite{Robson:2018ifk}:
\begin{equation}
    S_\text{n}(\nu) \simeq \frac{10}{3}\left(P_\text{OMS}(\nu)+\frac{2P_\text{acc}(\nu)}{(2\pi \nu)^4}\left[1+\cos^2\left(\frac{\nu}{\nu_*}\right)\right]\right)\times \left[1 + 0.6\left(\frac{\nu}{\nu_*}\right)^2\right]
\label{eq:lisa_sensitivity}
\end{equation}
where, for sake of completeness, $\nu_* = 19.09$ mHz is a characteristic frequency, while the optical metrology and test mass acceleration noises are respectively given by \cite{Babak:2021mhe}
\begin{align}
      P_\text{OMS}(\nu) &= 3.6\cdot10^{-41}\text{ Hz}^{-1}\times\left(1+\left[\frac{2\text{ mHz}}{\nu}\right]^4\right) \\
    P_\text{acc}(\nu) &= 1.44\cdot10^{-48} \text{ Hz}^{-1}\times\Big(1+\Big(\frac{0.4\rm{mHz}}{\nu}\Big)^2\Big)\times\left(1+\left[\frac{\nu}{8\text{ mHz}}\right]^4\right) 
    \nonumber \;.
\end{align}
This model assumes a LISA arm length $L = 2.5\times 10^9$ m.

A source is deemed loud when the inclination and polarization-averaged signal-to-noise ratio (SNR) $\chi$ exceeds a critical signal-to-noise ratio $\chi_c$, where for the DWDs, we take a fixed value $\chi_c = 8$. The SNR is computed as follows \cite{Robson:2018ifk,Ginat:2023eto}:
\begin{equation}
    \chi^2 = 4\int \dd \nu\, \frac{\overline{h}^2(\nu)}{S_\text{n}(\nu)}\,\Pi_T(\nu)
\end{equation}
where we have defined 
\begin{equation}
\overline{h}^2(\nu) \equiv \frac{\expval{\tilde{h}(\nu,\imath)\tilde{h}^*(\nu,\imath)}_\imath}{\mathcal{R}}=\frac{4}{5}\,{h_{0}}^2(\nu,r,\vxi) \;.
\end{equation}
Note that the parameter vector $\vxi$ does not include the inclination $\imath$, which is averaged out from this expression.
On using Eq. \eqref{limitwindow}, we obtain
\begin{equation}
\begin{split}
    \chi^2 &= \frac{16}{5}\, T \int \dd \nu \,\abs{\dv{\nu}{t}}\delta\big(\nu-\nu(\vr,\vxi)\big)\,\frac{{h_{0}}^2(\nu,r,\vxi)}{S_\text{n}(\nu)}\\
    &= \frac{16}{5}\, T \,\Bigg(\abs{\dv{\nu}{t}}\,\frac{{h_{0}}^2(\nu,r,\vxi)}{S_\text{n}(\nu)}\Bigg)_{\nu = \nu(\vr,\vxi)} \;.
\label{eq:snrcomp}
\end{split}
\end{equation}
The LISA observation-time window $T$ is set to $T=5$ years from here on.

\begin{figure}[h!]
    \centering
    \includegraphics[width=0.7\textwidth]{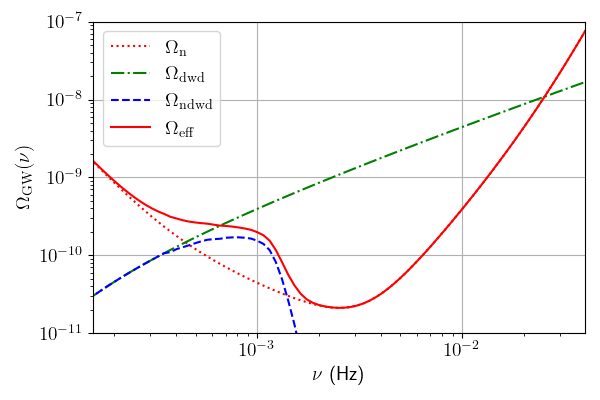}
    \caption{The dotted-dashed (green) curve is the GW energy density $\Omega_\text{dwd}(\nu)$ of the unsubtracted foreground of Galactic DWDs. The dashed (blue) curve represents the energy density of the confusion noise $\Omega_\text{ndwd}(\nu)$ left over after the subtraction of loud DWD sources (blue dashed) in a $T=5$ yr LISA observational run with $\chi_c=8$. The dotted (red) curve shows the LISA instrumental noise, while the solid (red) curve is the effective LISA noise $\Omega_\text{eff} = \Omega_\text{ndwd}+\Omega_\text{n}$ we use for the search of Galactic PBH binaries.}
    \label{fig:dwdresults}
\end{figure}

The subtraction of loud DWDs above the LISA noise can be conveniently captured by the function $\Theta_\text{n}$ defined as 
\begin{align}
    \Theta_\text{n}(\vr,\vxi_\text{dwd}) &\equiv H\big(\chi_c -\chi(\vr,\vxi_\text{dwd})\big) \nonumber\;,
\end{align}
where $H(x)$ is the Heaviside step function ($H(x)=1$ for $x\geq 0$, and zero otherwise).
The energy density of the confusion noise of unresolved DWDs follows from the insertion of $\Theta_\text{n}$ into the expression Eq.~\eqref{omega_def} of the unsubtracted GW energy density, i.e.
\begin{align}
\label{omega_bright}
     \Omega_\text{ndwd}(\nu) 
     &= \frac{16\pi^2N_\text{dwd}\nu^3}{15{H_0}^2} \int \dd\vr\,\dd \mathfrak{m}_1\dd \mathfrak{m}_2 \,\phi_\text{dwd}(\vr,\vxi_\text{dwd})   \,\abs{\dv{a_\text{dwd}}{t}} \\
     &\qquad \times {h_0}^2(\nu,r,\vxi_\text{dwd})\,\Theta_\text{n}(\vr,\vxi_\text{dwd})\bigg|_{a_\text{dwd}=a_\text{dwd}(\nu,\mathfrak{m}_1,\mathfrak{m}_2)} \nonumber \;.
\end{align}
The effect of the subtraction of loud DWD sources on the energy density of the Galactic GW background is shown in Fig. \ref{fig:dwdresults}. For convenience, we have defined a dimensionless noise spectral power
\begin{equation}
    \Omega(\nu) = \frac{4\pi^2}{3H_0^2}\nu^3 S(\nu) 
\end{equation}
analogous to the energy density of the actual GW signals. Furthermore, we have assumed an observational time window of $T=5$ yr.
The confusion noise of unresolved DWD GW sources represented by the dashed blue curve is consistent with other estimates from the literature \cite{Georgousi:2022uyt,Lin:2022huh,Boileau:2021sni,Nissanke:2012eh,Korol:2021pun,Wagg:2021cst,Farmer:2003pa,Wu:2023bwd}. The solid red curve is the effective noise level which we shall adopt for our assessment of the detectability of Galactic PBH binaries. 

\subsection{The GW background of Galactic PBH binaries}
\label{sec:PBHresults}

In contrast to DWDs, Galactic PBH binaries are highly eccentric and thus emit GW radiation across a large number of harmonic frequencies $\nu_n\equiv n\nu_0$ (where $n\geq 1$ is any positive integer) of the fundamental (orbital) frequency $\nu_0$. We refer the reader to Appendix \S\ref{sec:spectrald} for a brief review of the spectral distribution of eccentric GW sources.

Computing the exact, time-domain waveform of a GW signal produced by an eccentric compact binary remains a difficult task (see, e.g., {\cite{Huerta:2016rwp,Knee:2022hth} for recent attempts). Therefore, we take another route and introduce an effective, orbit-average strain $h_{0,n}$ following the methodology of \cite{Barack:2003fp, Enoki:2006kj, Huerta:2015pva, Wagg:2021cst,Bondani:2023hmk,Raidal:2024odr}. To define $h_{0,n}$, let us recast the GW energy spectrum into the form given in \cite{phinney:2001},
\begin{equation}
    \label{eq:phinney}
    \Omega_\text{gw}(\nu) = \frac{1}{\rho_c}\int\! \dd\vr\,\dd\vxi \, n(\vr,\vxi)\,\frac{1}{4\pi r^2}\,\frac{\dd E_\text{gw}}{\dd\ln\nu} \;.
\end{equation}
Here, $\rho_c=3 c^2 H_0^2/8\pi G$ is the present-day critical density, $n(\vr,\vxi)$ is the number density of sources per unit distance along the line of sight with location $\vr$ and parameter $\vxi$. In addition, $\dd E_\text{gw}/\dd \nu$ is the energy radiated by a single source in the frequency interval $[\nu,\nu+\dd \nu]$. The factor of $(4\pi r^2)^{-1}$ arises because $E_\text{gw}$ is a luminosity (integrated over solid angles) rather than a flux. It is given by
\begin{equation}
    \frac{\dd E_\text{gw}}{\dd\ln\nu} = \frac{\dd E}{\dd t}\nu\frac{\dd t}{\dd\nu} = P\,\frac{\nu}{\dot\nu} = P \,\bigg\lvert \frac{T}{\dot T}\bigg\lvert \;,
\end{equation}
where $P$ is the power radiated in GWs and $T$ is the orbital period. 
First, consider circular sources for which only the $n=2$ harmonics contribute. We have 
\begin{equation}
    \frac{\dd E_\text{gw}}{\dd\ln\nu} = \frac{\dd E_{e=0}}{\dd\ln\nu_2} = P_{e=0} \, \bigg\lvert \frac{T}{\dot T}\bigg\lvert_{e=0}= \frac{\pi^{2/3}}{3} \frac{(G M_c)^{5/3}}{G} \nu_2^{2/3} \;.
\end{equation}
Here, $E_{e=0}$ and $P_{e=0}$ are the GW energy and power radiated by a circular source at the quadrupolar frequency $\nu_2(\vxi)$, respectively. Substituting this result into Eq.~(\ref{eq:phinney}), we arrive  at
\begin{equation}
    \label{eq:ph1}
    \Omega_\text{gw}(\nu) = \frac{16\pi^2\nu^3}{15{H_0}^2}\int\! \dd\vr\,\dd\vxi \,c\,n(\vr,\vxi)\,{h_0}^2(\nu_2,r,\vxi) \;.
\end{equation}  
Comparing this expression with Eq.~(\ref{omega_def}), we can identify an event rate per line of sight distance, and per unit spatial and parameter volume according to 
\begin{equation}
    \label{eq:ph2}
    c\, n(\vr,\vxi) = N\,\phi(\vr,\vxi)\,\abs{\dv{\nu}{t}}\delta\big(\nu-\nu_2\big)
\end{equation}
Consider now eccentric binaries which radiate GWs in all the harmonics $\nu_n(\vxi)=n\nu_0(\vxi)$. Denoting the GW energy and power radiated in the $n$th harmonic by $E_n(e)$ and $P_n(e)$, respectively, $\dd E_n(e)/\dd\ln \nu$ is
\begin{equation}
    \frac{\dd E_n(e)}{\dd\ln \nu_n} = P_n(e) \, \bigg\lvert \frac{T}{\dot T}\bigg\lvert_{e\ne 0} = \frac{\pi^{2/3}}{3} \frac{(G M_c)^{5/3}}{G}\, \nu_n^{2/3} \times \left(\frac{2}{n}\right)^{2/3} \frac{g(n,e)}{f(e)} \;.
\end{equation}
    The factor of $1/f(e)$ arises through $|T/\dot T|_{e\ne 0}$ whereas $g(n,e)$ emerges from the power $P_n$ radiated in the $n$th harmonics. The quantities $P_{e=0}$, $P_n(e)$, $g(n,e)$ and $f(e)$ are all defined in Appendix~\S\ref{sec:spectrald}. Eqs.~(\ref{eq:ph1}) and (\ref{eq:ph2}) show that the generalization of $\Omega_\text{gw}(\nu)$ to eccentric sources is 
\begin{equation}
    \label{eq:omegagwwithe}
    \Omega_\text{gw}(\nu) = \frac{16\pi^2N\nu^3}{15{H_0}^2}\sum_{n=1}^\infty \int\! \dd\vr\,\dd\vxi \,\phi(\vr,\vxi)\,\abs{\dv{\nu}{t}}\,\delta(\nu-\nu_n)\, h_{0,n}^2(\nu_n,r,\vxi) \;.
\end{equation}    
We have introduced an effective orbit-average strain 
\begin{align}
    \label{eq:effectivestrain}
    h_{0,n}(\nu_n,\vr,\vxi) &\equiv h_0(\nu_n,\vr,\vxi) \sqrt{\gamma_n(e)} \\
    \gamma_n(e) &\equiv \left(\frac{2}{n}\right)^{2/3}\frac{g(n,e)}{f(e)} \nonumber \;,
\end{align}
which captures the orbit-averaged GW emission by eccentric sources. For large eccentricities $1-e \ll 1$ the function $g(n,e)$ peaks at harmonic number $n_\text{max} \sim (1-e^2)^{-3/2}\sim j^{-3}$ \cite{Wen:2002km}. For a Galactic PBH binary with a typical reduced angular momentum value of $j \sim 10^{-2} - 10^{-1}$, the peak of GW emission occurs at harmonic numbers $n_\text{max} \sim 10^3 - 10^6$, that is, at a frequency orders of magnitude larger than $\nu_0$. 

Specializing Eq.~(\ref{eq:omegagwwithe}) to the Galactic population of PBH binaries, we have
\begin{equation}
\begin{split}
     \Omega_\text{pbh}(\nu) = \frac{16\pi^2N_0\nu^3}{15{H_0}^2} \sum_{n=0}^\infty\int\! \dd\vr\,\dd\vxi_\text{pbh} \, \phi_\text{pbh}(\vr,\vxi_\text{pbh})\,\abs{\dv{\nu}{t}}\delta\big(\nu-\nu_n(\vxi_\text{pbh})\big)\,{h_0}^2(\nu,\vr,\vxi_\text{pbh})\gamma_n(e_0)
\label{omega_pbh}
\end{split}
\end{equation}
The vector of binary parameters $\boldsymbol{\xi}_\text{pbh}$ comprises the reduced angular momentum $j_0$ and the semi-major axis $a_0$ of the binary. There is no integral over a PBH mass due to the assumption of a monochromatic mass function. Taking into account Galactic disruption, the distribution function $\phi_\text{pbh}$ is given as:
\begin{equation}
\phi_\text{pbh}(\vr,\vxi_\text{pbh}) = \phi_\text{h,dm}(\vr)\, \phi_\text{pbh}(a_0,j_0)\,\Pi_\text{d}(a_0,\vr)\\
\end{equation}
To simplify the calculation of Eq.~(\ref{omega_pbh}), we take advantage of the large expected $n_\text{peak}$ and approximate the sum of harmonic numbers by a continuous integral, ignoring the remainder term of the Euler-MacLaurin formula. Furthermore, in complete analogy with the calculation of $\Omega_\text{gw}$ for Galactic DWDs, we get rid of the Dirac delta by integrating over the semi-major axis $a_0$ to obtain
\begin{equation}
\begin{split}
     \Omega_\text{pbh}(\nu) = \frac{16\pi^2N_0\nu^3}{15{H_0}^2} \int\! \dd\vr\,\dd j_0 \,\dd n \,\phi_\text{pbh}(\vr,\vxi_\text{pbh})\,\abs{\dv{a_0}{t}}\,{h_0}^2(\nu,\vr,\vxi_\text{pbh})\gamma_n(e_0)\Bigg|_{a_0=a_{0}(\nu,n,\vxi_\text{pbh})}
\end{split}
\end{equation}
Note that $a_0$ is now a function of $\nu$ and $n$ because the orbital frequency satisfies $\nu_0=\nu_n/n\equiv \nu/n$. In the particular case of equal-mass binaries with non-zero eccentricity, we have
\begin{equation}
    a_0(\nu,n,m) = \left(\frac{n^2Gm \text{ M}_\odot}{2\pi^2}\right)^{1/3}\nu^{-2/3} \;.
\end{equation}
In addition,
\begin{equation}
    \abs{\dv{a_0}{t}} = \frac{64\pi^2}{5} \frac{G^2(m\text{ M}_\odot)^2}{c^5} \left(\frac{2}{n}\right)^2\nu^2 f(e_0)\;,
\end{equation}
which cancels the multiplicative factor of $1/f(e_0)$ present in $\gamma_n(e_0)$.

The results of this computation are shown in Fig.\ref{fig:pbhbresults}, where the various colored curves show $\Omega_\text{pbh}(\nu)$ for different choices of $m$ and $f$ as quoted on the figure. We have applied a Gaussian kernel of rms width 0.06 dex in log space to reduce the numerical noise. The (black) solid curve indicates the effective LISA noise level including the DWD confusion noise, as calculated in Section \S\ref{sec:DWDnoise}. For $m=0.1$ and 1, which bracket the DWD mass range, the peak frequency of $\Omega_\text{pbh}(\nu)$ lies around the minimum of the effective noise even though the bulk of Galactic PBH binaries have semi-major axes much larger than $1\AU$. The reason is the presence of extremely high harmonics induced by highly eccentric motions, which shift the peak of the GW emission frequency to much higher frequencies relative to wide but circular binaries. 

We also observe that lower mass PBHs exhibit a peak at larger frequencies due to their lower typical separation (see Fig. \ref{fig:finalpdf}). Furthermore, increasing $f$ by one order of magnitude increases $\Omega_\text{GW}$ by approximately two orders of magnitude, such that $\Omega_\text{pbh}(\nu)$ for a model with $f=0.1$ and $m=1$ (not shown on this figure because of the PBH constraints in this model) would be above the effective LISA noise, when binary disruption by single PBHs is neglected (see Footnote \ref{fn:3body}). For reference, we also indicate in Fig.~\ref{fig:pbhbresults} the expected sensitivity curve of the Deci-hertz Interferometer Gravitational-Wave Observatory (DECIGO) \cite{Kawamura:2020pcg} and the Big Bang Observer (BBO) \cite{Harry:2006fi} in their most simple design as reported by \cite{Yagi:2011wg}\footnote{We do not include stochastic backgrounds in the DECIGO/BBO sensitivity curves of Fig. \ref{fig:pbhbresults} because of uncertainties in the instrumental design and in the modelling of the backgrounds. Possibilities for backgrounds in DECIGO/BBO include e.g. extragalactic DWDs \cite{maselli20}, primordial GWs and compact object inspirals \cite{Kawamura:2019jqt}.} These experiments should be able to detect the background produced by Galactic subsolar mass PBHs down to mass $m=0.01$ if $f\simeq 0.01$. 

\begin{figure}[h!]
    \centering
    \includegraphics[width=1\textwidth]{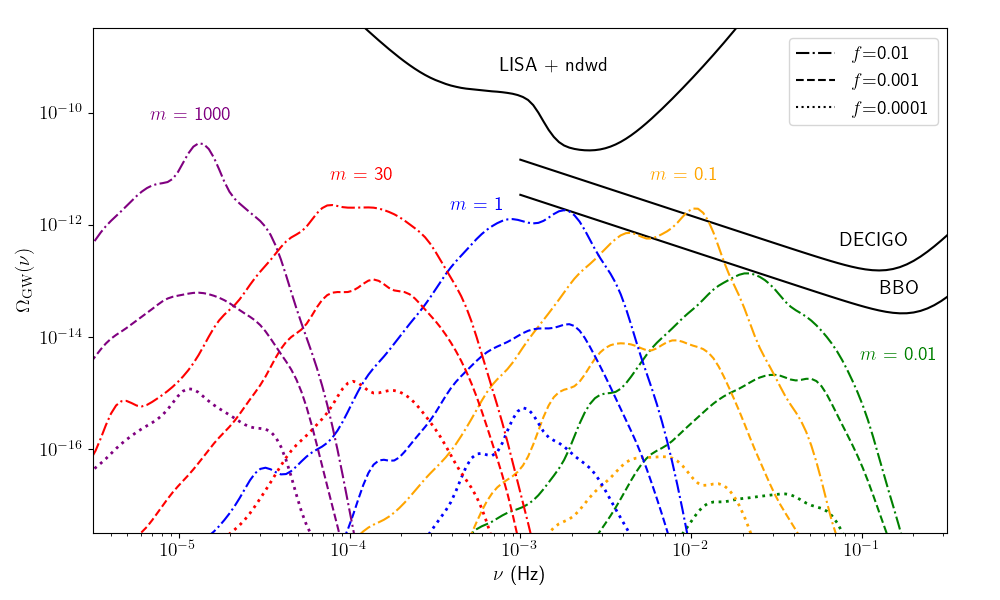}
    \caption{The GW energy density $\Omega_\text{pbh}(\nu)$ of a Galactic population of PBH binaries as a function of the observed GW frequency $\nu$. Results are shown for different choices of $m$ and $f$. The solid black curve represents the effective LISA noise level, which includes the confusion noise of unresolved Galactic DWDs. The sensitivity curves of the DECIGO and BBO experiments are taken from \cite{Yagi:2011wg}.}
    \label{fig:pbhbresults}
\end{figure}

\subsection{Loud Galactic PBH Binaries}
\label{loudpbhbs}

The identification of loud PBH binaries in the Milky Way is analogous to the extraction of loud Galactic DWDs performed above, expect that we adopt a total noise power spectral density $S_\text{eff} = S_\text{n} + S_\text{ndwd}$. Accounting for the non-zero eccentricity of the Galactic PBH binaries, the SNR $\chi_\text{pbh}$ is given by 
\begin{equation}
\begin{split}
    {\chi_\text{pbh}}^2 
    &= \frac{16}{5}\, T\sum_{n=0}^\infty  \,\Bigg(\abs{\dv{\nu}{t}}\,\frac{{h_{0,n}}^2(\nu,\vr,\vxi)}{S_\text{eff}(\nu)}\Bigg)_{\nu = \nu_n(\vr,\vxi)}\\
    &\simeq \frac{16}{5}\, T\int \dd n  \,\Bigg(\abs{\dv{\nu}{t}}\,\frac{{h_{0,n}}^2(\nu,\vr,\vxi)}{S_\text{eff}(\nu)}\Bigg)_{\nu = \nu_n(\vr,\vxi)} 
\end{split}
\end{equation}
where $T$ is, again, the LISA observational time window which we set to $T=5$ yr hereafter.
Note that $h_{0,n}$ is an orbit-average strain and so is $\chi_\text{pbh}$. However, one should bear in mind that, for highly eccentric orbits, most of the GW power is emitted around pericenter passage in a burst of GWs \cite[see][for a recent discussion]{xuan/etal:2024,Xuan:2023azh}. For the whole population of Galactic PBH binaries, this burst emission averages out but, for the small expected number of loud sources, it may play a role depending on their eccentricities. We will come back to this point once the loud sources have been identified.

A PBH binary is deemed loud if it satisfies $\chi_\text{pbh}(\vr,\vxi_\text{pbh}) > \chi_c$, where, just as for the DWD foreground, $\chi_c$ is the detection threshold. A reasonable detection threshold is $\chi_c\sim 7 - 8$, whereas $\chi_c=1$ corresponds to a typical fluctuation. In Fig.~\ref{fig:snrparamspace}, regions of the $a_0$-$j_0$ parameter subspace for which the condition $\chi_\text{pbh}(\vr,\vxi_\text{pbh}) =\chi_c$, is fulfilled with $\chi_c=1$ or $\chi_c=8$ are shown as the dashed and solid curves, respectively. 
Loud sources mostly emerge from the tail of the distribution produced by Galactic PBH binaries having significantly hardened through GW emission. Therefore, they are hardly affected by disruption processes taking place in the MW.
For illustration, we have assumed that the loud sources reside in the MW halo (where binary disruption is certainly negligible, see~\S\ref{sec:disruption}) and are located at a distance $r=100$ pc and $10\kpc$ from the Sun. Note that, while the present-day distribution of PBH binaries in the MW halo (shown as the shaded area) depends on the PBH fraction, contours of constant SNR are independent of $f$.

\begin{figure}[h!]
    \centering
    \includegraphics[width=1.1\textwidth]{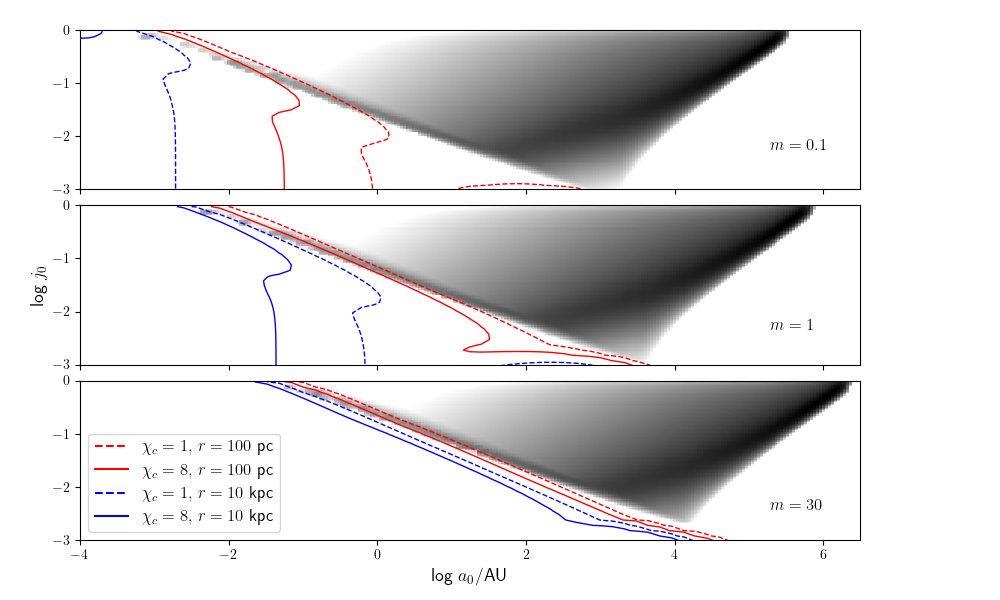}
    \caption{The dashed and solid curves indicate the locus in the $a_0$ - $j_0$ plane for which $\chi_\text{pbh}(\vr,\vxi_\text{pbh}) = \chi_c$. The loud sources with SNR exceeding $\chi_c$ are located leftward of the various curves, which assume Galactic PBH binaries residing in the MW halo at a distance $r=100$ pc and $10\kpc$ from the Sun. The dark shaded area represents the present-day distribution of MW halo PBH binaries for a PBH fraction $f=0.01$. Note, however, that contours of constant SNR are independent of $f$. Results are shown for 3 different values of $m$ (panels from top to bottom)}
    \label{fig:snrparamspace}
\end{figure}

The expected number $N_\text{loud}$  of loud Galactic PBH binaries detected by LISA after 5 years of observation is  
\begin{equation}
    N_\text{loud} = N_0 \int \dd\vr\,\dd\vxi_\text{pbh} \phi_\text{phb}(\vr,\vxi_\text{pbh})\tilde{\Theta}_\text{t}(\vr,\vxi_\text{pbh})
\end{equation}
where the ``clipping'' function $\tilde{\Theta}_\text{t}$ is
\begin{equation}
\tilde{\Theta}_\text{t}(\vr,\vxi_{pbh}) \equiv H(\chi_\text{pbh}(d,\vxi_\text{pbh})-\chi_c) \;.
\end{equation}
In practice, we do not take into account disruption in the computation of $N_\text{loud}$ since, as stated above, it is irrelevant for the loud sources. Furthermore, in the computation of the SNR, we use the sky-averaged effective noise level $S_\text{eff}$ given by the sum of the sky-averaged DWD noise and the sky-averaged LISA instrumental noise. This approximation is conservative for PBH binaries in the MW halo, where the foreground of unresolved DWDs is reduced relative to the disk or its sky-averaged value~\footnote{Incorporating the precise angular dependence of the LISA instrumental noise levels is beyond the scope of this paper.}. 

\begin{figure}[h!]
    \centering
    \includegraphics[width=1\textwidth]{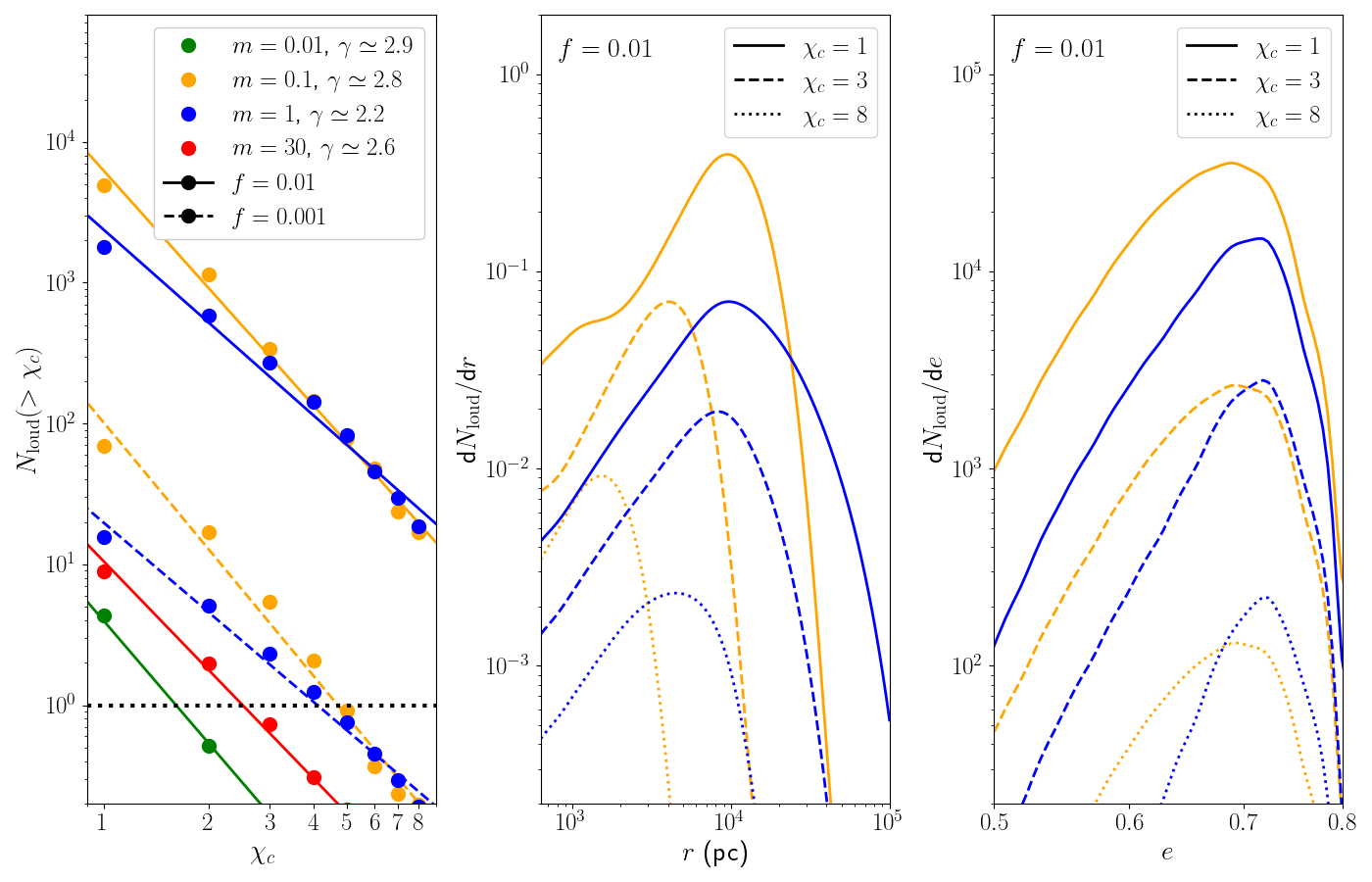}
    \caption{{\it Left}: the cumulative number $N_\text{loud}(>\chi_c)$ of loud Galactic PBH binaries in 5 years of LISA observations as a function of the critical SNR $\chi_c$. The lines represent power law fits to the cumulative counts. The best-fit power law index $\gamma$ is quoted in the insert. {\it Middle:} Distribution of the loud PBH binaries as a function of their distance $r$ from the solar system for three values of $\chi_c$. Results are shown for $f=0.01$ and $m=0.1$, 1 following the color scheme adopted in the left panel. Notice the correlation between $\gamma$ and the relative amount of loud sources at large distances (see text) {\it Right:} Same as the middle panel, but showing the ecccentricity distribution of the loud PBH binaries. Note that the area below the distributions is proportional to $N_\text{loud}(>\chi_c)$.}
    \label{fig:Nloud}
\end{figure}

The expected number of loud sources are shown in the left panel of Fig.~\ref{fig:Nloud} for different choices of $\chi_c$.
A conservative detection threshold of $\chi_c = 8$ leads to $\sim 20$ loud sources in 5 years LISA data for $m = 0.1 , 1$ and $f=0.01$. For the other parameter choices, $\chi_c$ must be decreased for the cumulative count $N_\text{loud}(>\chi_c
)$ of loud PBH binaries to exceed unity 
(For $m=1000$ not shown here, $N_\text{loud}$ is far below unity even for $\chi_c  = 1$).
Overall, the latter increases with decreasing $\chi_c$ following a power law $N_\text{loud}(>\chi_c) {\propto \chi_c}^{-\gamma}$. The power law index would be $\gamma=3$ for a homogeneously distributed population since the detection rate then scales linearly with spatial volume. This power-law is also consistent with a strain probability density function $P(h) \sim h^{-4}$ at large $h$ \cite{Ginat:2019aed}, which gives $\mathrm{d}N_\text{loud}/\mathrm{d}\chi_{c} \propto \chi_c^{-4}$ \cite{Ginat:2023eto}. 
Deviations from $\gamma=3$ are caused by the inhomogeneity of the spatial distribution of Galactic PBH binaries (which traces the MW halo). More precisely, cumulative counts with power law indices further away from $\gamma = 3$ correspond to distance distributions $dN_\text{loud}/dr$ reaching larger separations (where the effects of the inhomogeneous spatial PBH profile are larger). This is quite apparent in the middle panel of Fig.~\ref{fig:Nloud}, where the loud PBH binaries in the $m=1$ model are distributed further away than in the $m=0.1$ model. The increase in the characteristic distance of loud sources thus leads to a decrease in spatial homogeneity and results in a larger deviation from $\gamma = 3$. This correlation holds also for the models not shown in Fig.~\ref{fig:Nloud}.
The right panel of Fig.~\ref{fig:Nloud} shows that, unlike the bulk of the Galactic PBH binaries, the loud sources are only mildly eccentric, with typical eccentricities $\sim 0.7$. As a result, the GW signal is distributed over a relatively small number of harmonics, which makes the source extraction easier. Even though the number of loud sources is small, their mild eccentricity also justifies (a posteriori) the fact that we have ignored the orbital phase in our calculation of the GW signal (i.e. we have not taken into account possible burst-like GW events as discussed in~\S\ref{sec:gwhardening}).

The spatial distribution (especially the directional information) of the loud sources could help to discriminate between primordial BH and compact stellar remnants in addition to the (possibly subsolar) mass (and tidal effects, see \cite{crescimbeni/etal:2024}). Loud PBH sources are the hard binaries located in the ``merging'' tail of the ($\log a$-$\log j$) plane (see Fig. \ref{fig:snrparamspace}), with time to coalescence mostly concentrated in the range $t_\text{coal}\sim 10^2 - 10^5 $ yr (with a peak around $10^3-10^4$ yr) and semi-major axis $a_0 \lesssim 1\AU$. As a result, disruption is irrelevant for them, and they trace the MW halo (modelled as a NFW profile). The identification of loud PBH binaries is easiest away from the MW bulge and disk, where the contamination by DWDs and other stellar binaries is lowest.

\section{Conclusions}
\label{sec:conclusions}

The existence of stellar-mass PBHs is allowed in multi-component dark-matter models provided that they do not exceed a percent of the total dark-matter energy density. In these scenarios, close enough PBH pairs can decouple from the Hubble flow to form PBH binaries, which trace the adiabatic mode. As cosmic structures form, these binaries are accreted onto dark-matter halos and can masquerade as stellar BH binaries. In the MW halo, they contribute to the Galactic GW background, which should be found by future GW experiments. 

In this paper, we have computed the GW signal produced by such a hypothetical population of Galactic PBH binaries assuming a monochromatic PBH mass function and a broad range of PBH mass. 
For this purpose, we have modeled the evolution of PBH binaries from their formation in the early Universe until the present epoch, including the hardening of PBH binaries through GW emission and the disruption of soft binaries by stellar encounters in the MW. Stellar disruption depletes the MW disk and bulge of most of the PBH binaries.
In addition to its anisotropy, the present-day population of Galactic PBH binaries is characterized by high values of orbital eccentricity, which distinguishes them from the Galactic DWD population. These large eccentricities increase the amplitude and characteristic frequency of the GW signal produced by Galactic PBH binaries. For a PBH mass in the range $M_\text{pbh}\sim 0.1 - 10 M_\odot$, the GW background produced by Galactic PBH binaries peaks at the millihertz frequencies probed by LISA. Still, the peak of the GW energy density is at least 1-2 orders of magnitude below the effective LISA noise level (obtained upon subtracting resolved Galactic DWDs from the signal) even for PBH a fraction $f=0.01$. However, proposed experiments like DECIGO and BBO should resolve such a background if it consists of subsolar PBHs with $M_\text{pbh}\sim 0.01 - 1 M_\odot$. 

We have estimated the ability of LISA to detect loud Galactic PBH binaries. The cumulative SNR distribution of the loudest sources is generally a power-law function of the critical SNR $\chi_c$ used to define loud sources. For a PBH fraction $f\lesssim 0.01$ consistent with current limits and a subsolar mass in the range $M_\text{pbh}\sim 0.1 - 1 M_\odot$, the cumulative number of loud sources with $\chi_c =  8$ in 5 years LISA data is $\sim 20$.
The loud sources are hard PBH binaries tracing the MW PDM profile and, owing to the steepness of the latter, are preferentially located near the center of the MW halo. They are characterized by GW coalescence times of order $10^3 - 10^4$ yr, and eccentricities $e\sim 0.7$ not as high as those of the total population of MW PBH binaries. Therefore, they might masquerade as an eccentric Galactic binary BH population of stellar origin \cite[see e.g.][]{DOrazio:2018jnv,Fang/Thompson/Hirata:2019,Gupte:2024jfe,xuan/etal:2024,Xuan:2023azh}) if their mass is at least a few $M_\odot$. In this case, the spatial location could help discriminate between primordial and stellar binary BHs. To accurately quantify prospects on PBH binary identification using directional information, detailed knowledge of LISA's orientation and angular sensitivity is required. We plan to explore this in future work. 
We furthermore note again that we have not taken into account "late-time" PBH binaries forming through or involving dynamical capture mechanisms. This later population will resemble stellar populations more and might influence Galactic stellar binary populations as well \citep[e.g.][]{Bhalla:2024jbu}. 

In the evolution of PBH binaries from their formation until the present epoch, we have taken into account a limited number of physical mechanisms. In particular, we have neglected the possible clustering of PBHs \cite{alihaimoud:2018, Desjacques:2018wuu,ballesteros/etal:2018, Stasenko:2024pzd, Bringmann:2018mxj, Belotsky:2018wph, inman/alihaimoud:2019} and the mass accretion of PBHs \cite{ricotti:2007,ricotti/etal:2008,deluca/etal:2020,rice/zhang:2017}. Clustering is deemed not important for small values of $\fpbh$ \cite{deluca/etal:2020a}, mass accretion onto highly eccentric binaries is not understood well enough to be properly implemented\cite{Ali_Ha_moud_2017}. Furthermore, PBH may be dressed with a cloud of PDM of approximately the same mass as the PBH at the time of MR equality, which will affect the distributions \cite{Ali_Ha_moud_2017}. Other relics from the early Universe could also surround them \citep[e.g.][]{Nishikawa:2017chy}.
Finally, we have ignored hardening by 3-body encounters, which may speed up the binary evolution, and even replace one of the PBHs by an astrophysical object. Refs.~\cite{Raidal_2019,Ali_Ha_moud_2017} find that this effect hardly affects the present-day PBH merger rate, but, for the whole population of Galactic PBH binaries (which includes many soft binaries), it may be more significant. Further work should also include late-time effects like dynamical capture, and ascertain the extent to which stellar populations and Galactic GW backgrounds of stellar origins are affected by the presence of field PBHs.

\acknowledgments

We are grateful to Gabriele Franciolini, Robert Reischke, Toni Riotto, and Sam Young for helpful discussions. We acknowledge support from the Israel Science Foundation (grant no. 2562/20). Y.B.G. was partly supported by a Leverhulme Trust International Professorship Grant to S. L. Sondhi (No. LIP-2020-014), and partly supported by the Simons Foundation via a Simons Investigator Award to A. A. Schekochihin.

\newpage

\appendix

\section{Gravitational waves from eccentric compact binaries}
\label{sec:eccentricgwb}


Eccentricity plays a substantial role in the evolution of a population of PBHs as emphasized here and in previous works \cite{Ali_Ha_moud_2017, Raidal_2019, Bondani:2023hmk}.

In this Appendix, we discuss the backreaction effect of eccentric GW radiation on the orbital parameter evolution needed in Section \ref{sec:gwhardening}. In addition, we briefly review the vacuum GW emission of an eccentric Keplerian binary used in Section \ref{sec:PBHresults} to model the GW signal of a Galactic population of PBH binaries. Most of this discussion is based on the seminal works of \cite{Peters:1963ux,Peters:1964qza}, which is conveniently summarized in \cite{Maggiore:2007ulw}. Note that, although we assume a monochromatic PBH mass function throughout this paper, we write below expressions valid for a generic unequal-mass binary with a total mass $M$ and reduced mass $\mu$. 

\subsection{Orbital evolution}
\label{sec:orbitalevolution}

The gravitational wave emission is entirely determined by the mass of the binary components, and the semi-major axis $a$ and eccentricity $e$ of the binary orbit. If the binary hardens through GW emission solely, these orbital parameters evolve via the system of coupled ordinary differential equations (ODEs) in Eq.~\eqref{coupled_original} (which involves two distinct timescales $t_a$ and $t_e$ characterizing the evolution of $a$ and $e$, respectively). The latter can be combined into
\begin{equation}
    \dv{a}{e} = \frac{12}{19}a\frac{1+(73/24)e^2+(39/96)e^4}{e(1-e^2)[1+(121/304)e^2]}\;.
\end{equation}
This can be integrated analytically to give
\begin{equation}
    a(e) = a_* \frac{\mathcal{G}(e)}{\mathcal{G}(e_*)} \;,
    \label{eq:ae}
\end{equation}
where the function $\mathcal{G}(e)$ is defined as
\begin{equation}
    \mathcal{G}(e) \equiv \frac{e^{12/19}}{1-e^2}\left(1+\frac{121}{304}e^2\right)^{870/2299} \;.
    \label{Gfunc}
\end{equation}
Eq.~(\ref{eq:ae}) can be used to calculate the time to coalescence $t_\text{coal}$ of a binary from the relation $\int\dd t=\int\dd e\,\dv{t}{e}$, in which $\dv{e}{t}$ is given by the second ODE of Eq.~\eqref{coupled_original}. The dependence of $t_\text{coal}$ on the initial eccentricity $a_*$ is encoded in the function
\begin{equation}
    \label{eq:Fe}
    F(e_*) \equiv \frac{48}{19}\frac{1}{\mathcal{G}^4(e_*)}\int_0^{e_*}\dd e \; \frac{\mathcal{G}^4(e)\big(1-e^2\big)^{5/2}}{e\big(1+\frac{121}{304}e^2\big)}
\end{equation}
Finding a numerical solution to the system Eq.~\eqref{coupled_original} for a general time $t<t_\text{coal}$ is computationally challenging due to the vastly different timescales $\tau_a$ and $\tau_e$ when $e$ is close to unity. However, it is possible to find a semi-numerical solution for $e(\tau,a_*,e_*)$ as explained below.

\begin{figure}
\centering
\includegraphics[width=0.6\textwidth]{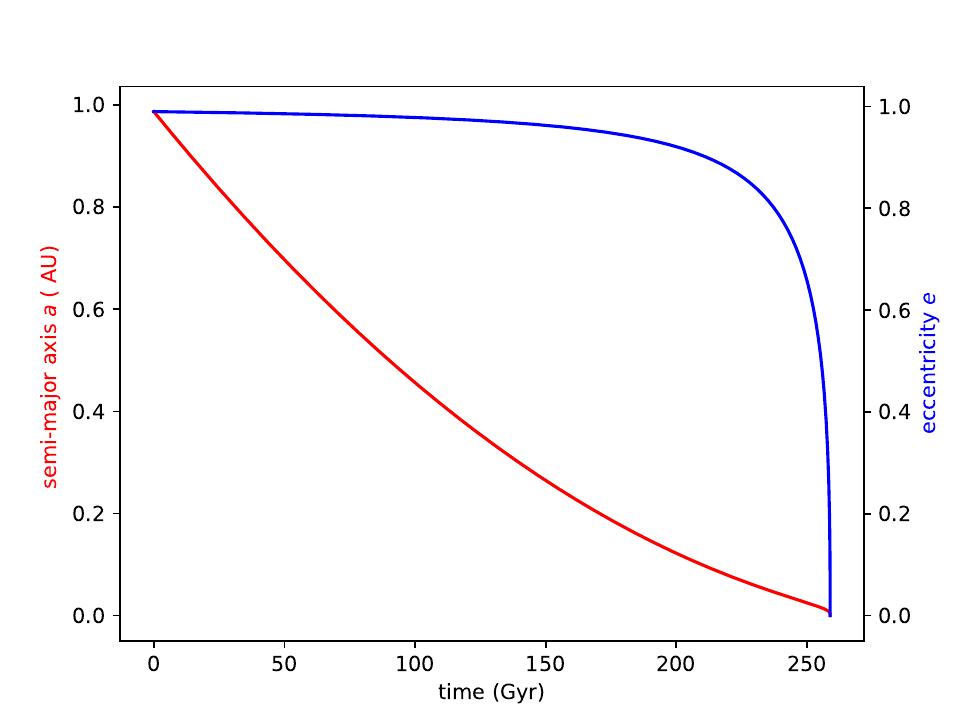}
\caption{Orbital evolution for a binary with initial semi-major axis $a_i=1$ AU and eccentricity $e_i = 0.99$ obtained from the solution outlined in \S\ref{sec:appelsolution}. Note that the decay timescale of the semi-major axis is much shorter than the circularization timescale.}
\label{fig:example_sol}
\end{figure}

\subsection{Semi-numerical solution to the eccentric evolution}
\label{sec:appelsolution}

The solution $e(\tau)$ can be conveniently spelled out in terms of the dimensionless variables $\tilde{a} \equiv a/R_*$ and $\tau \equiv ct/R_*$, where
\begin{equation}
    {R_*}^3 \equiv \frac{4G^3\mu M^2}{c^6}
\end{equation}
becomes the Schwarzschild radius $R_* =2GM/c^2$ for equal mass PBH binaries. The second line of Eq.~\eqref{coupled_original} becomes
\begin{equation}
    \dv{e}{\tau} = -\frac{76}{15}\frac{e}{{\tilde{a}}^4(1-e^2)^{5/2}}\left(1+\frac{121}{304}e^2\right) \;.
\end{equation}
Substituting $a(e)$ given by Eq.~\eqref{eq:ae} into this equation, we can write down an expression for the time $\tau$ elapsed since the formation of the PBH binary,
\begin{equation}
\begin{split}
    \tau &= \int_0^\tau\dd \tau' = \int_{e_i}^e \Big(\dv{e'}{\tau'}\Big)^{-1}\dd e' = \frac{15}{76}\int_e^{e_*} \dd e' \frac{{\tilde{a}}^4(1-e'^2)^{5/2}}{e'[1+(121/304){e'}^2]}\\
    &=\frac{15}{76}\frac{{\tilde{a_*}}^4}{{\mathcal{G}(e_*)}^4}\int_e^{e_*}\dd e' \frac{{e'}^{29/19}}{\big(1-{e'}^2\big)^{3/2}}\Big(1+\frac{121}{304}{e'}^2\Big)^{1181/2299} \;.
\end{split}
\label{integral}
\end{equation}
This integral admits a solution in terms of Appell hypergeometric functions of two variables, which read
\begin{equation}
    F_1(\alpha,\beta_1,\beta_2,\gamma,x,y) = \sum_{m=0}^\infty \sum_{n=0}^\infty \frac{(\alpha)_{m+n}(\beta_1)_m(\beta_2)_n}{m!n!(\gamma)_{m+n}}x^my^n
\end{equation}
where
\begin{equation}
    (z)_p = \prod_{k=0}^{p-1}[z-k]
\end{equation}
is the Pochhammer symbol. It is convenient to introduce the function 
\begin{equation}
    \label{eq:Ie}
    I(e) \equiv \frac{e^{10/19}}{3648}\Big(I_0(e)-3648 A_1(e) -893 e^2 A_2(e)\Big)
\end{equation}
where
\begin{equation}
    \begin{split}
        A_1(e) &\equiv F_1\left(\frac{5}{19},\frac{1}{2},\frac{1118}{2299},\frac{24}{19},e^2,-\frac{121}{304}e^2\right)\\
        A_2(e) &\equiv F_1\left(\frac{24}{19},\frac{1}{2},\frac{1118}{2299},\frac{43}{19},e^2,-\frac{121}{304}e^2\right)\\
        I_0 (e) &\equiv \frac{24\times 2^{2173/2299}\times 19^{1118/2299}}{\sqrt{1-e^2}}\Big(304+121e^2\Big)^{1181/2299} \;.
    \end{split}
\label{eq:Iappel}
\end{equation}

\begin{figure}
\centering
\includegraphics[width=0.6\textwidth]{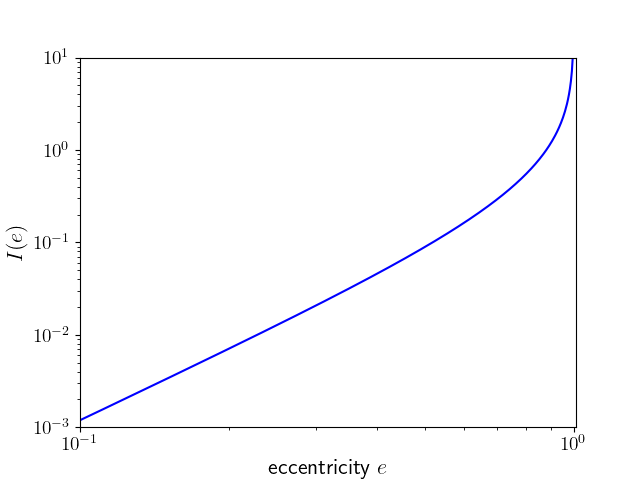}
\caption{The function $I(e)$ defined in Eq. \eqref{eq:Ie}. $I(e)$ diverges in the limit $e \rightarrow 1$.}
\label{fig:Ie}
\end{figure}

The function $I(e)$ is shown in Fig. \ref{fig:Ie}. The solution to Eq.~\eqref{integral} can eventually be recast into the form
\begin{equation}
    \tau(e,a_*,e_*) = \frac{15}{76}\frac{{\tilde{a_*}}^4}{{\mathcal{G}(e_*)}^4} \Big(I(e_*)-I(e)\Big) \;. 
\end{equation}
The only piece of the calculation left over is the inversion of this expression to find $e(\tau,a_*,e_*)$, which can be carried out using standard numerical routines. The solution for $a(\tau,a_*,e_*)$ then follows from Eq.~\eqref{eq:ae}, that is, from the knowledge of $e(\tau)$ and the initial conditions $(a_*,e_*)$. Fig.~\ref{fig:example_sol} shows a sample solution for $(a_*,e_*)=(1\AU,0.99)$. For very eccentric orbits, circularization through GW emission is a subdominant effect until the very last stages of the evolution.

\subsection{Spectral Distribution}
\label{sec:spectrald}

Unlike circular binaries emitting GWs at a frequency $\nu=2\nu_o$, twice the orbital frequency $\nu_0$, eccentric binaries emit GWs at all the harmonics $\nu_n=n\nu_0$ of the orbital frequency. In addition, the orbital eccentricity boosts the radiated GW power by a factor of $f(e)$ given by
\begin{equation}
    f(e) \equiv \frac{1}{\left(1-e^2\right)^{7/2}}\left(1+\frac{73}{24}e^2+\frac{37}{96}e^4\right) \;.
    \label{f-func}
\end{equation}
As a result, the total, orbit-averaged power radiated in GWs becomes
\begin{equation}
    P_{e\ne 0} = \frac{32G^4\mu^2M^3}{5c^5a^5}f(e) \equiv P_{e=0}\, f(e)
    \label{power}
\end{equation}
for $e>0$. For large eccentricities $1-e\ll1$, $P_{e\ne 0}$ can be enhanced by orders of magnitude. This power is emitted at the discrete frequencies $\nu_n$ and is distributed among them according to
\begin{equation}
    P_{e\ne 0} = \sum_{n=1}^\infty P_n(e)
\end{equation}
where the $n$-th harmonic contributes a power $P_n(e)$ given by
\begin{equation}
    P_n(e) = P_{e=0}\, g(n,e) \;.
\end{equation}
The auxiliary functions
\begin{equation}
\begin{split}
    g(n,e) = \frac{n^4}{32}\Bigg(&\Big[J_{n-2}(ne)-2eJ_{n-1}(ne)+\frac{2}{n}J_n(ne)+2eJ_{n+1}(ne)-J_{n+2}(ne)\Big]^2\\&+(1-e^2)\Big[J_{n-2}(ne)-2J_n(ne)+J_{n+2}(ne)\Big]^2+\frac{4}{3n^2}\Big[J_n(ne)\Big]^2\Bigg) \;,
\end{split}
\end{equation}
where $J_n$ is the Bessel function of the first kind, satisfy the completeness relation
\begin{equation}
    \sum_{n=1}^\infty g(n,e) = f(e) \;.
\end{equation}
They are helpful for the calculation of the GW energy density produced by a distribution of eccentric binaries (see \S\ref{sec:PBHresults}).

\section{Dynamical Friction in the Milky Way Halo}
\label{sec:DF}

Galactic PBH binaries also harden through Dynamical Friction (DF) with the surrounding PDM distribution. The DF timescale can be estimated with the Chandrasekhar formula \cite{chandrasekhar:1943}~\footnote{Although dynamical friction is the gravitational deceleration produced by a nonlocal density wake, local approximations provide a good estimate of the DF force for inhomogeneous systems \cite[see, e.g.,][]{tremaine/weinberg:1984,weinberg:1989,eytan/etal:2024}.}. Assuming that the PBH binary moves in a circular orbit with velocity $v^2 \simeq GM_\text{pbh}/a$ relative to its center-of-mass, the relation $t_\text{DF}\sim vM_\text{pbh}/F_\text{DF}$, where $F_\text{DF}$ is the magnitude of the DF force, yields
\begin{equation}
    t_\text{DF}(r) \sim 10^8 \text{ Gyr } \sqrt{m}\left(\frac{a}{\text{AU}}\right)^{-3/2} \left(\frac{\rho_\text{PDM}(r)}{10^{-2} \text{ M}_\odot \text{pc}^{-3}}\right)^{-1}I_\text{DF}^{-1} \;.
\end{equation}
Here, $r$ is the distance from the Galactic Center and $\rho_\text{PDM}(r)$ is the local PDM density distributed according to the NFW profile Eq. \ref{haloprof}. Furthermore, $I_\text{DF}$ is a (velocity-dependent) dimensionless friction coefficient, which we take to be unity. 

\begin{figure}
    \includegraphics[width=0.5\textwidth]{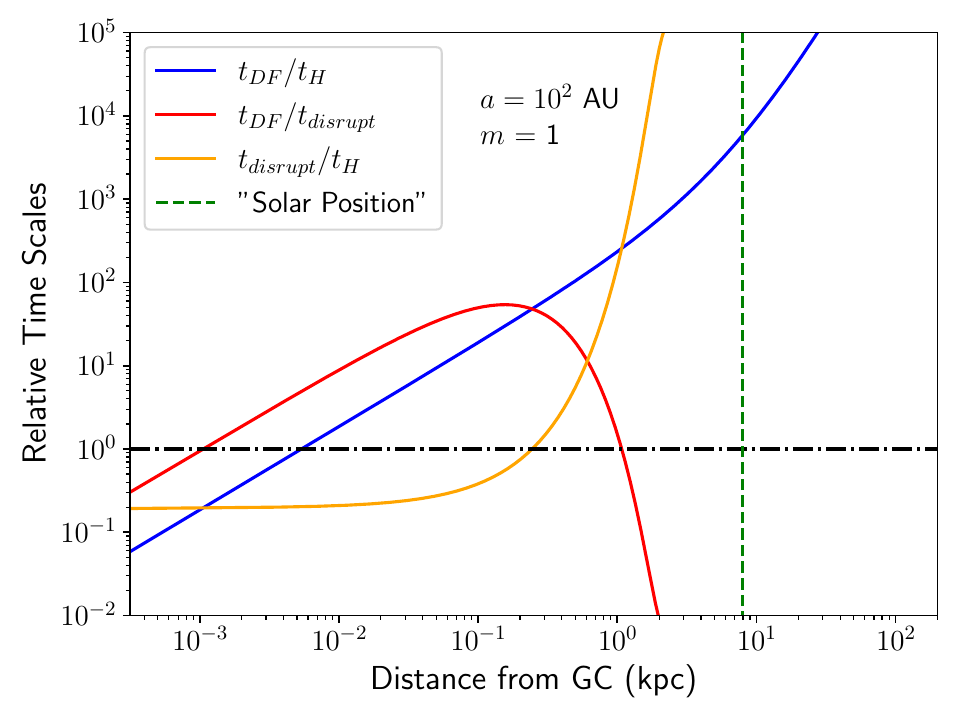}
    \includegraphics[width=0.5\textwidth]{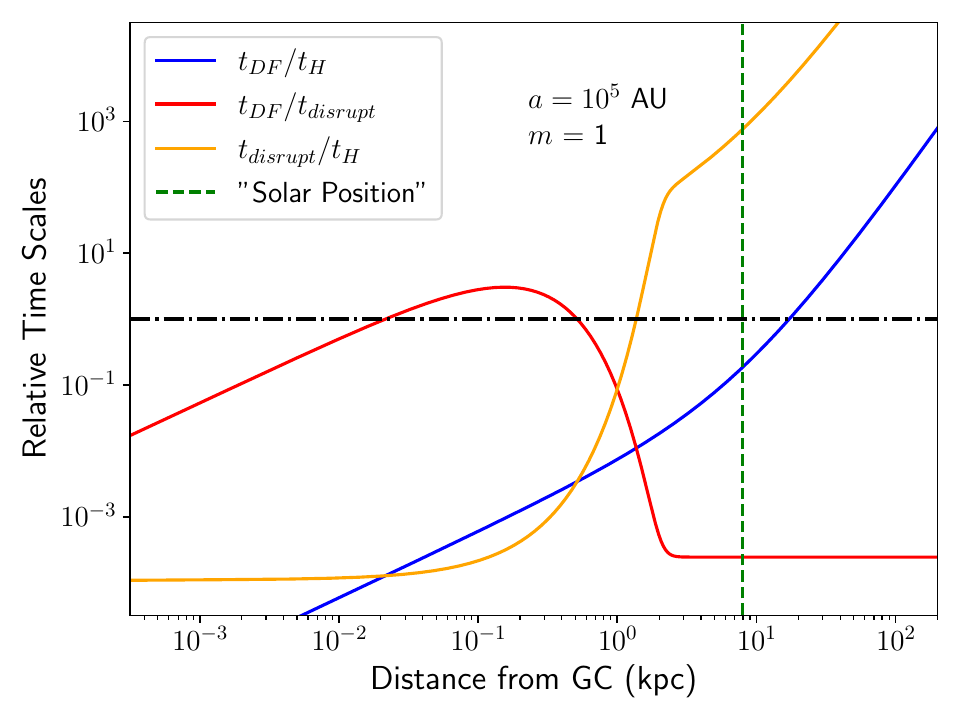}
  \caption{A comparison between the timescales characterizing binary disruption ($t_{disrupt}$) and dynamical friction ($t_{DF}$) as well as the present-day Hubble time $t_H$. The various ratios are shown as a function of the separation $r$ from the GC for a soft (\textit{left panel}) and hard (\textit{right}) PBH binary with $m=1$. The vertical (green) dotted line marks the radial distance of our solar system. }
  \label{fig:dftime}
\end{figure}

Fig.~\ref{fig:dftime} compares the PDM-induced DF timescale with the disruption timescale (defined in Section \ref{sec:disruption}) and the (present-day) Hubble time for a PBH binary of solar mass ($m=1$) at different locations in the stellar halo. For relatively hard PBH binaries (left panel), disruption by MW halo stars or single PBHs dominates deep inside the MW virial radius. For $r\gtrsim 1\kpc$, DF is the dominant process but the characteristic timescale $t_\text{DF}$ is too long to lead to a significant loss of orbital energy within a Hubble time. For softer binaries (right panel), DF is dynamically relevant for separations $1\lesssim r\lesssim 20\kpc$ from the GC. However, given the limited region of the full parameter space this corresponds to, we decided to neglect DF in our evolution model. Note also that, for the comparison with the disruption timescale, we selected a slice through the MW without the stellar disks (i.e. it is approximately orthogonal to the stellar disks). Including the stellar disks increases the effect of disruption and even better justifies the discarding of DF.
 
\bibliographystyle{JHEP}
\bibliography{Bibliography}

\end{document}